\documentclass[aps,prd,showpacs,nofootinbib,preprint]{revtex4-1}
\usepackage{hyperref}
\usepackage{latexsym}
\usepackage{epsf}
\usepackage{epsfig}
\usepackage{amssymb, amsmath, amsfonts, amsxtra, amsthm}%-----
\usepackage{bm}%----------bold math
\usepackage{dcolumn}%----------Align table columns on decimal point
\usepackage{array}
\usepackage{tabularx}
\usepackage{enumerate}
\usepackage{hhline}
\usepackage{subfigure}
\usepackage{color}
\usepackage{graphics}

\definecolor{linkcolor}{RGB}{41, 127, 255}
\hypersetup{
	pdfstartview={XYZ},
	colorlinks,
	allcolors=linkcolor,
	bookmarksopen
}

\begin{document}
%opening
%\title{Higgs as heavy-lifted physics during inflation}
\title{\boldmath Higgs as heavy-lifted physics during inflation}
%\date{\today}

\date{\today}

\author{Yi-Peng Wu$^a$}\email[Electronic address: ]{ypwu@resceu.s.u-tokyo.ac.jp}
\affiliation{${}^a$Research Center for the Early Universe (RESCEU),	Graduate School of Science, The University of Tokyo,	Tokyo 113-0033, Japan}

%\author[a]{Yi-Peng Wu}

%\author{Yi-Peng Wu$^a$}\email[Electronic address: ]{ypwu@resceu.s.u-tokyo.ac.jp}
%\affiliation{${}^a$Research Center for the Early Universe (RESCEU),	Graduate School of Science, The University of Tokyo,	Tokyo 113-0033, Japan}
%\author{Louis Yang$^b$}
%\affiliation{${}^b$Kavli Institute for the Physics and Mathematics of the 
%	universe (Kavli IPMU), \\WPI, UTIAS, The University of Tokyo, 5-1-5 Kashiwanoha, Kashiwa 277-8583, Japan}	

%\affiliation[a]{Research Center for the Early Universe (RESCEU), Graduate School of Science\\The University of Tokyo, Tokyo 113-0033, Japan}

%\emailAdd{ypwu@resceu.s.u-tokyo.ac.jp}

%\keywords{Cosmology of Theories beyond the SM, Effective Field Theories}

%\arxivnumber{1812.10654}

	\begin{abstract}
		Signals of heavy particle production during inflation are encoded as non-analytic momentum scaling in primordial non-Gaussianity. These non-analytic signatures can be sourced by Standard Model particles with a modified Higgs scale uplifted by the slow-roll dynamics of inflation. We show that such a lifting mechanism becomes more efficient with the presence of a strong Higgs-inflaton mixing, where the Higgs mass scale is further increased by a small speed of sound in the effective theory of inflation. As a primary step towards detecting new particles in the cosmological collider program, non-Gaussianity due to heavy Higgs production in the strong-mixing regime can act as important background signals to be tested by future cosmological surveys.
	\end{abstract}

\preprint{RESCEU-17/18}

\maketitle
%\flushbottom

%\preprint{RESCEU-17/18}

\section{Introduction}

Heavy fields are expected to arise from the UV completion in most of the single-field inflationary scenarios \cite{Baumann:2014nda}. 
%if the full theory admits a stable flat direction for the slow-roll inflation.
These heavy states, while stablizing a flat direction for the slow-roll inflation, could have typical masses around the size of the Hubble parameter $H$ during inflation, especially when the UV physics is associated with the supersymmetry breaking or is constructed in the framework of supergravity \cite{Copeland:1994vg,Baumann:2011nk,Yamaguchi:2011kg,Assassi:2013gxa}.
Although the energy scale of the UV physics seems too high to be probed directly by the ground-based particle accelerators, these heavy fields can leave imprints in the primordial non-Gaussianity once they are spontaneously produced by the high energy quantum fluctuations during inflation \cite{Chen:2009we,Chen:2009zp,Baumann:2011nk,Noumi:2012vr,Gong:2013sma,Arkani-Hamed:2015bza,Kehagias:2015jha,Kehagias:2017cym,Franciolini:2017ktv,Meerburg:2016zdz,Lee:2016vti,Dimastrogiovanni:2015pla,Schmidt:2015xka,MoradinezhadDizgah:2018ssw,Wang:2018tbf,Saito:2018omt,Kumar:2018jxz,Goon:2018fyu}.
Imprints of heavy particle production include characteristic signals from the non-analytic scaling of the momentum restricted by the symmetry of inflation \cite{Arkani-Hamed:2015bza,Kehagias:2015jha,Arkani-Hamed:2018kmz} (see also \cite{Chen:2009we,Chen:2009zp,Baumann:2011nk,Noumi:2012vr}). These non-analytic signals are not covered by any local effective operator that is converted from integrating out the heavy degree of freedom. 
%through their quantum interference with inflaton (or the curvature perturbation $\zeta$).  
%In particular, the squeezed limit of the bispectrum is a very promising window for detecting signatures of heavy particles.
%Heavy fields may play a fundamental role in stablizing the flat direction for a slow-roll cosmic inflation.

Recently, attention has been paid to the possible non-analytic signals sourced from those particles with accessible masses to ground-based accelerators, yet their mass scales may be lifted above $H$ during the primordial epoch of cosmic inflation \cite{Kumar:2017ecc,Chen:2016hrz,Chen:2016uwp}. For instance, the Standard Model (SM) Higgs boson $h$, if not the inflaton $\phi$ itself, is known to gain a mass $m_h \sim H$ much larger than the electroweak scale in a de Sitter background via a non-minimal coupling to gravity or non-trivial interactions with inflaton. With an unbroken gauge symmetry, the heavy Higgs field (and of course all the SM gauge fields) can enter the $\zeta$-correlation functions through loop interactions, leaving meaningful non-Gaussianity for observations only in limited parameter space \cite{Chen:2016hrz,Chen:2016uwp}.

The production of heavy SM particles becomes more efficient if the gauge symmetry is (partially) broken where Higgs spontaneously obtains a non-zero vacuum expectation value (VEV) during inflation. The symmetry breaking is manifested by a tachyonic Higgs mass from the wrong-sign non-minimal coupling term or the higher-dimension Higgs-inflaton interactions \cite{Kumar:2017ecc,He:2018gyf,Chen:2018xck}. By assuming that higher dimensional Higgs-inflaton interactions are a series of well-controlled low-energy effective operators, perhaps reduced from grand unified theories, Kumar and Sundrum \cite{Kumar:2017ecc} concluded that non-analytic signals in bispectra are always dominated by the heavy Higgs production processes.

%In this work, we relax constraints on the model parameters asked by a well-defined effective theory in \cite{Kumar:2017ecc}. Instead, we consider higher dimensional interactions as exact operators, allowing Higgs and inflaton to form a strongly coupled system. These exact operators could arise from the symmetry breaking of an extended group structure of the SM.
However, the cutoff scale for not entering the dynamical region of heavy fields becomes non-trivial if the dispersion relation of the system is non-linearly modified \cite{Baumann:2011su,Gwyn:2012mw,Ashoorioon:2018uey,Gwyn:2014doa}. The main focus for the current study is to show that the mass spectrum of SM particles can be non-trivially lifted by a modified dispersion relation due to the presence of strong mixing terms with inflaton.
%In a general setup without making any heavy degree of freedom dynamical-excited, one can include non-trivial momentum dependence in the low-energy effective operators \cite{Gwyn:2012mw}. The interesting parameter space uncovered by momentum-dependent operators becomes the main focus for the current study.
 In this work, we use the Higgs-inflaton system as an example to show that cubic or higher-order perturbations in the Lagrangian can be well-defined, even if quadratic perturbations are strongly mixed. One can identify a modified mass scale corresponding to the strong mixing, below which Higgs acts as a heavy degree of freedom. Integrating out the heavy Higgs field in the strong-mixing regime shall result in an effective speed of sound for inflaton, which captues the leading analytic (local) contribution of the heavy field to the primordial spectra \cite{Baumann:2011su,Iyer:2017qzw,Gwyn:2012mw,Gwyn:2014doa}.

Due to the implicit symmetries of the inflationary background, one can expand mode functions into power-laws $\sim\eta^\Delta$ of the conformal time in the late-time limit $\vert\eta\vert \rightarrow 0$, where $\Delta$ characterizes the spatial dimension of an operator residing on the boundary of future infinity ($\eta = 0 $) \cite{Arkani-Hamed:2018kmz} (see also \cite{Chen:2009zp,Baumann:2011nk,Noumi:2012vr}). 
For the simplest non-Gaussian observable led by the three-point function $\langle \zeta^3 \rangle$, the late-time expansion implies that the squeezed limit exhibits the scaling as
\begin{align}
\frac{\langle\zeta^3\rangle}{\langle\zeta^2\rangle_S \langle\zeta^2\rangle_L} \sim \sum_i w_i \left(\frac{k_L}{k_S}\right)^{\Delta_i},
\end{align}
where $k_L/k_S \ll 1$ and $k_L$ is a long-wavelength mode that exits the horizon much earlier than the short-wavelength mode $k_S$. $w_i$ are coefficients depending on the mass of the heavy Higgs field. 
If the SM gauge symmetry is unbroken during inflation, one expects the non-analytic scaling takes \cite{Arkani-Hamed:2015bza}
\begin{align}\label{scaling L_weak}
\Delta_{\rm non-analytic} = 3\pm 2\, i L_h,  \qquad L_h = \sqrt{\frac{m_h^2}{H^2}-\frac{9}{4}},
\end{align}
since the Higgs field must enters the correlation functions as a pair. For $m_h/H > 3/2$, the non-analytic contribution creates oscillatory featurs in the bispectrum, which can be taken as the signature of heavy particle production during inflation.

In this work, we consider the heavy Higgs production by virtue of a large mixing with inflaton that spontaneously breaks the gauge symmetry and give a non-zero VEV to the Higgs field. Our expectation for the non-analytic scaling in a strong-mixing system is thus modified as
\begin{align}\label{scaling L_strong}
\Delta_{\rm non-analytic} = \frac{3}{2}\pm \, i L_h,  \qquad L_h = \sqrt{\frac{m_h^2}{H^2 c_\phi^2}-\frac{9}{4}},
\end{align}
where the half-integer from single Higgs production is allowed due to the broken gauge symmetry, and $c_\phi^2$ is the effective speed of sound for inflaton which accounts for the local effect via integrating out the Higgs field. Comparing \eqref{scaling L_weak} and \eqref{scaling L_strong}, one can identify a modified mass scale $\mu_h \equiv m_h/c_\phi$ for the strong-mixing system, which recovers the non-local extension of the effective-field-theory approach for inflation \cite{Gwyn:2012mw}.  
The modified late-time scaling behavior opens a new parameter space for observing Higgs (or other gauge fields) in primordial non-Gaussianity. To be specific, we show that the oscillatory feature in the bispectrum is generated when $m_h/(c_\phi H) > 3/2$, instead of $m_h/H > 3/2$.

This paper is organized as the follows. In Section~\ref{Sec. EFT}, we review the framework for the heavy-lifting mechanism in the effective theory approach. We consider a model in Section~\ref{Sec. energy_scales} as an example to show that the Higgs mass scale can be heavy-lifted by a strong mixing with inflaton, resulting in a modified dispersion relation without breaking the perturbativity of the loop expansion. Imprints involved with the heavy Higgs production in the non-Gaussian correlation functions of $\zeta$ are studied in Section~\ref{Sec. imprints}. Finally, conclusions are given in Section~\ref{Sec. Conclusion}.

\section{The effective theory for heavy-lifting}\label{Sec. EFT}
In this section we clarify the framework in which the effective theories for the heavy-lifting mechanism is built. We first recall the most general effective action of the Goldstone boson, $\pi$, corresponding to the spontaneously broken time-translation symmetry with one extra scalar field, $\sigma$, as a massive degree of freedom during inflation \cite{Cheung:2007st,Noumi:2012vr,Senatore:2010wk}. This formalism (sometimes called the $\pi$-$\sigma$ model \cite{Baumann:2011su}) assumes no specific fundamental physics but includes all possible terms with respect to the time-dependent spatial diffeomorphisms.
\footnote{The $\pi$-$\sigma$ model is also known as quasi-single field inflation \cite{Chen:2009we,Chen:2009zp} for the mass $m_\sigma$ of the $\sigma$ field is near or larger than the Hubble parameter $H$. 
%In this work, we use the $\pi$-$\sigma$ model for discussion that in general includes the case with $m_\sigma \ll H$.
}

We then specify the class of theory proposed in \cite{Kumar:2017ecc} for realizing a spontaneous symmetry breaking of the Higgs potential during inflation. One can directly write down an effective Lagrangian of the Higgs field $h$ as a low-energy expression of some unified theory with inflaton $\phi$ at very high energy scales. Separating Higgs and inflaton as different particles (namely non-Higgs inflation) generically makes it easier to obtain observable signatures in non-Gaussianity. To make sure such an effective expansion is well-controlled several conditions have been imposed in \cite{Kumar:2017ecc} to keep the Higgs-inflaton interactions weakly coupled. Although it is in general very difficult to identify the exact couplings in the unified theory, $\phi$ could be naturally light if it is associated with the ``pion'' of the spontaneous symmetry breaking of the full theory and $h$ would obtain a mass of the size of $H$. This possibility will be discussed in the next section. Note that in the effective theory of the $\phi$-$h$ model we always assume the slow-roll inflation paradigm that exhibits a well-defined homogeneous time evolution, and thus it can be cast into as a special case of the $\pi$-$\sigma$ model.
\subsection{The cosmological Goldstone boson}
We follow the standard precedure to construct the most general effective action of inflation \cite{Cheung:2007st,Baumann:2014nda}. The first step is to write down all time-dependent operators that preserve spatial diffeomorphisms in the unitary gauge where there are only metric fluctuations. At leading order the terms with the metric perturbation $\delta g^{00} = g^{00}+1$ are   
\begin{align}\label{action_pi}
S_\pi = \int d^4x \sqrt{-g} \left[ \frac{M_p^2}{2} R  + M_p^2\dot{H} g^{00} - M_p^2 (\dot{H}+3H^2) \right.\\\nonumber
 \left. \qquad +\frac{ M_2^4(t) }{2!}(\delta g^{00})^2 + \frac{ M_3^4(t)}{3!} (\delta g^{00})^3 + \cdots \right],
\end{align}
where operators with the extrinsic curvature perturbation $\delta K_{\mu\nu}$ and with higher derivatives are not shown here since the action \eqref{action_pi} is sufficient for our purpose. 

Following \cite{Noumi:2012vr} with the same setup as for $S_\pi$, the effective action for an extra scalar field $\sigma$ during inflation is given by
\begin{align}\label{action_sigma}
S_\sigma &= \int d^4x \sqrt{-g} \left[ -\alpha_1(t)g^{\mu\nu}\partial_\mu\sigma\partial_\nu\sigma + \alpha_2(t) (\partial^0 \sigma)^2 
- \alpha_3(t) \sigma^2 + \alpha_4(t)\sigma\partial^0\sigma + \cdots \right], \\
\label{action_pi_sigma}
S_{\pi\sigma} &= \int d^4x \sqrt{-g} \left[
\beta_1(t) \delta g^{00}\sigma + \beta_2(t)\delta g^{00} \partial^0\sigma + \beta_3(t)\partial^0\sigma  \right. \nonumber\\
&\qquad\qquad\qquad \left. -(\dot{\beta}_3(t) +3H\beta_3(t))\sigma+ \cdots \right],
\end{align}
where we introduce the interactions between $\pi$ and $\sigma$ fields through $S_{\pi\sigma} $. We refer the combination of actions \eqref{action_pi}, \eqref{action_sigma} and \eqref{action_pi_sigma} as the $\pi$-$\sigma$ model.
 
To simplify the formalism for our later convenience, we want to rewrite the $\pi$-$\sigma$ model in the decoupling limit with gravity. Following the discussion in \cite{Baumann:2011su,Baumann:2014nda}, the decoupling limit with gravitational fluctuations is realized by taking $M_p \rightarrow \infty$, $\dot{H} \rightarrow 0$ with $M_p^2 \dot{H}$ fixed, which is a trick similar to making the ``pion'' and the ``sigma field'' to be decoupled in the well-known linear sigma model.

For this purpose we shall restore the full gauge-invariance to the actions by virtue of the replacements $t \rightarrow t + \pi$,
$g^{00} \rightarrow g^{00} + 2\partial^0 \pi + \partial_\mu\pi \partial^\mu \pi$, and
$\partial^0 \sigma \rightarrow \partial^0 \sigma + \partial_\mu\pi \partial^\mu \sigma $. One can solve gravitational fluctuations in terms of the standard ADM variables (and up to first order in $\pi$ and $\sigma$ is enough for our case), as performed in \cite{Noumi:2012vr}. 
Putting solutions of the ADM variables back into the actions, such as \eqref{action_pi} and \eqref{action_pi_sigma}, and taking $H^2/M_p^2 \rightarrow 0$ with $\dot{H}/H^2 \rightarrow 0$, we find the simplified actions as
\begin{align}\label{action_pi_dec}
S_\pi \rightarrow \int d^4x \sqrt{-g} \, \frac{M_p^2 \dot{H}}{c_\pi^2}\left[- \left(\dot{\pi}^2 -c_\pi^2 \frac{(\partial_i\pi)^2}{a^2}\right)-(1-c_\pi^2) \left(\dot{\pi}^3 -\dot{\pi} \frac{(\partial_i\pi)^2}{a^2}\right)\right] +\cdots,
\end{align}
where $c_\pi^{-2} = 1- 2M_2^4/(M_p^2 \dot{H})$, and
\begin{align}\label{action_pi_sigma_dec}
S_{\pi\sigma} \rightarrow \int d^4x \sqrt{-g} \, \left[(-2\beta_1 +\dot{\beta}_3)\dot{\pi}\sigma + (2\beta_2 - \beta_3)\dot{\pi}\dot{\sigma} 
+ \beta_3 \frac{\partial_i\pi\partial_i\sigma}{a^2} 
%-3\beta_3 \dot{H} \pi \sigma 
+ \cdots\right].
\end{align}
We can see that \eqref{action_pi_dec} and \eqref{action_pi_sigma_dec} capture nothing but the leading terms of the $\phi$-$h$ model in the following discussion.

\subsection{The Higgs boson during inflation}
While the $\pi$-$\sigma$ model contains every operator that is invariant under the spatial transformation $x^i \mapsto x^i +\xi^i(t, \textbf{x})$, it is also our desire to start with a high energy theory that specifies the explicit Lagrangian for inflaton, Higgs, or all the SM gauge fields. By specifying the Lagrangian of interest we can pin down a large number of free parameters to be tested by observations. 

In this work, we will assume that heavy states corresponding to physics much higher than the energy scale of inflation have been integrated out, resulting in an effective low-energy Lagrangian with non-trivial interactions between Higgs and inflaton (putting gauge fields aside for the moment). Masses of SM particles may be lifted up to the size of the Hubble parameter during inflation due to these interactions. For example, the heavy-lifting of Higgs mass considered in \cite{Kumar:2017ecc} has a general Lagrangian of the form 
\begin{align}\label{L_phi_h}
\mathcal{L}= \mathcal{L}_{\rm sr}(\phi) -\xi R \Phi_H^{\dagger} \Phi_H  -\vert D_\mu\Phi_H\vert^2  - \lambda(\Phi_H^\dagger \Phi_H)^2 + \mathcal{L}_{\phi h}, 
\end{align}
where $\Phi_H  = (0, h)^T \, /\sqrt{2}$ is the Higgs doublet in terms of the SM unitary gauge. The inflaton Lagrangian
\begin{align}
\mathcal{L}_{\rm sr}(\phi) = -\frac{1}{2} \left(\partial_\mu \phi\right)^2 - V(\phi),
\end{align}
collects all of the viable single-field inflation models that satisfie slow-roll conditions. 
The Higgs-inflaton interactions are parametrized by a series of dimensionless coefficients $c_i$ in
\begin{align}\label{L_pi_h}
\mathcal{L}_{\phi h} = \frac{c_1}{\Lambda_1} \partial_\mu\phi (\Phi_H^\dagger D^\mu \Phi_H) 
+ \frac{c_2}{\Lambda_2^2} (\partial_\mu\phi)^2 \Phi_H^\dagger \Phi_H 
+ \frac{c_3}{\Lambda_3^4}(\partial_\mu\phi)^2 \vert D_\mu\Phi_H\vert^2 + \cdots,
\end{align}
which can be taken as an effective expression of the full theory at the energy scale of inflation. As shown in \cite{Kumar:2017ecc}, these higher-dimensional couplings could arise due to the integrating out of some heavy particles that mediate between Higgs and inflaton separately only on very high energy scales.

A spontaneous symmetry breaking of the Higgs vacuum during inflation is the key for the heavy-lifting mechanism.
One of the possible origin to have a sizable VEV, namely $\langle h \rangle \equiv h_0 \sim H$, is the presence of the non-minimal coupling $\xi R h^2$ with $\xi < 0$ \cite{Kumar:2017ecc}. To gain a Higgs mass $m_h\sim H$ implies $\xi \sim \mathcal{O}(1)$. Alternatively one can consider a spontaneous symmetry breaking caused by an effective tachyonic mass term, for instance, due to the $c_2$-term in \eqref{L_pi_h}. This second possibility will be the main focus of the current study. 

Let ue perform a quick search to see the correspondence in between the $\phi$-$h$ coefficients $c_i$ and the parameters $\beta_i$ in the $\pi$-$\sigma$ model. We consider for simplicity a well-defined decomposition $h = h_0 +\delta h$ with $h_0$ a constant VEV. In the unitary gauge where $\delta\phi = 0$, the linear expansion of \eqref{L_pi_h} gives
\begin{align}
\mathcal{L}_{\phi h} = \frac{c_1}{2\Lambda_1} \dot{\phi}_0 \delta h \delta\dot{h} + \frac{c_1}{2\Lambda_1} \delta g^{00} \dot{\phi}_0 h_0 \delta\dot{h}
+\frac{c_1}{2\Lambda_1} \delta g^{0i} \dot{\phi}_0 h_0 \partial_i \delta h \nonumber\\
+ \frac{c_2}{2\Lambda_2^2} g^{00} \dot{\phi}_0^2 \delta h^2 +\frac{c_2}{\Lambda_2^2} \delta g^{00} \dot{\phi}_0^2 h_0 \delta h + \cdots,
\end{align}
where $\phi_0$ is the inflaton VEV during slow-roll. Since metric components in the decoupling limit with gravity read
\begin{align}
g^{00} \rightarrow -(1+\dot{\pi})^2 +\frac{(\partial_i\pi)^2}{a^2}, \quad \text{and} \quad g^{0i} \rightarrow \frac{\delta^{ij}\partial_j \pi}{a^2},
\end{align} 
the Lagrangian $\mathcal{L}_{\phi h}$ then introduces the $\pi$-$h$ interactions up to quadratic order as
\begin{align}
\mathcal{L}_{\phi h} \supset -\frac{c_1}{\Lambda_1}\dot{\phi}_0 h_0 \dot{\pi}\delta\dot{h} + \frac{c_1}{2\Lambda_1}\dot{\phi}_0 h_0 \frac{\partial_i \pi \partial_i \delta h}{a^2}
-2 \frac{c_2}{\Lambda_2^2} \dot{\phi}_0^2 h_0 \dot{\pi}\delta h.
\end{align}
Replacing $h$ by $\sigma$ we find from \eqref{action_pi_sigma} that
\begin{align}
\dot{\beta}_3 -2\beta_1 = -2 \frac{c_2}{\Lambda_2^2} \dot{\phi}_0^2 h_0, \quad 
2\beta_2 - \beta_3 = -\frac{c_1}{\Lambda_1}\dot{\phi}_0 h_0, \quad \text{and} \quad \beta_3 = \frac{c_1}{2\Lambda_1}\dot{\phi}_0 h_0 .
\end{align}
These results show how the $c_1$ and $c_2$ terms in the $\phi$-$h$ theory can be cast into the $\pi$-$\sigma$ model. One can also check the above
 correspondence in the flat-slicing gauge, where the gauge-invariant variable $\zeta = - H\pi = - H \delta\phi/\dot{\phi}_0$ implies $\pi = \delta\phi/\dot{\phi}_0$.

\section{Energy scales of heavy-lifting}\label{Sec. energy_scales}

In this section we consider the heavy-lifting scenario induced by $\phi$-$h$ interactions \eqref{L_pi_h}. We will use an example to demonstrate the spontaneous symmetry breaking of Higgs VEV,
and identify a characteristic energy scale $\mu_h$ below which the Higgs field $h$ represents a heavy degree of freedom. In the simplist case, $\mu_h = m_h$ is characterized by its mass scale and a heavy degree of freedom means that $h$ exhibits a constant dispersion relation $\omega \approx m_h$ for modes with physical wavenumbers $p = k/a(t) \ll m_h$. Thus if $H \gg 10^2$ GeV one would expect that the Higgs field is simply a light degree of freedom during inflation
since the SM value $m_h \approx 125$ GeV and the Higgs self-coupling $\lambda$ becomes small when running up to high energy scales \cite{Buttazzo:2013uya,Degrassi:2012ry,EliasMiro:2011aa}.

For a strongly-coupled $\phi$-$h$ system, we will examine the energy scale $\Lambda_p$ at which the higher-order perturbation expansion breaks down. In fact, we will show that the cutoff scale $\Lambda_p$ for the perturbative expansion does not rely on the $\Lambda_i$'s parametrized in \eqref{L_pi_h}. To simplify our discussion, we turn off the non-minimal coupling $\xi$ and assume a positive $\lambda \ll 1$.

\subsection{The Higgs-inflaton system}
We are interested in a system made by two fundamental scalars, which are the inflaton $\phi$ 
%that sources the symmetry breaking of the time-translation 
and the Higgs field $h$. To realize a spontaneous symmetry breaking during inflation, we consider as an example the classical Lagrangian of the form
\begin{align}\label{system}
\mathcal{L}= \mathcal{L}_{\rm sr}(\phi) -\Phi_H^\dagger \Phi_H \frac{\left(\partial_\mu \phi\right)^2}{\Lambda^2} -\vert D_\mu\Phi_H\vert^2  - \lambda(\Phi_H^\dagger \Phi_H)^2, 
\end{align}
where it can be taken as a special case of the $\phi$-$h$ theory \eqref{L_pi_h} with $\Lambda_2 =\Lambda$, $c_2 = 1$ and otherwise $c_i = 0$.
%$\Phi_H  = h \, e^{i \pi_i \sigma^i}/\sqrt{2}$ is the Higgs doublet in terms of the SM unitary gauge with would-be Goldstone components $\pi_i$ and $\sigma^i$ are the Pauli matrices. 
By taking the SM unitary gauge and omitting all SM gauge fields, the kinetic terms of the $\phi$-$h$ system read
\begin{align}\label{fieldspace1}
\mathcal{L} \supset -\frac{1}{2} \left(1 + \frac{h^2}{\Lambda^2}\right) \left(\partial_\mu \phi\right)^2 -\frac{1}{2} \left(\partial_\mu h\right)^2,
\end{align}
which represents a two-field limit of the multi-field inflation scenario based on the non-linear sigma model \cite{Achucarro:2010jv,Achucarro:2012sm,Elliston:2012ab}.
To justify \eqref{system} from the effective theory formulation, we shall consider an approximated shift symmetry in the inflaton sector so that non-derivative $\phi$-$h$ couplings are not appear (or highly suppressed) in the system. To our purpose, we should consider a fine cancellation of higher-order terms in the series of $h^2/\Lambda^2$, which allows us to explore the parametric region with $h > \Lambda$. On the other hand, higher-order terms in the series of $\partial\phi/\Lambda^2$ can still be controlled by the Naturalness condition introduced in Section \ref{subsec. naturalness}.
At a first glance, $\Lambda$ looks like a cutoff scale of the non-canonical kinetic interaction to make sure that \eqref{system} is a well-behaved low-energy effective Lagrangian. 
However, we will show that in the presence of a strong-mixing the conclusion is non-trivial. For example, a well-defined perturbative expansion for the $\phi$-$h$ system \eqref{system} is in fact independent of the ratio $h/\Lambda$.
%However, in general the theory \eqref{system} does not necessarily arise from such an effective formulation, as we will show that the perturbative expansion of the system is well-defined even in the limit of strong-coupling.
%A possible interpretation of the specific interaction in \eqref{system} is to consider a composite field defined as $\Phi_H^\prime \equiv \Phi_H e^{i \phi/\Lambda}$. In this definition the inflaton $\phi$ looks like a ``pion'' of the linear sigma model if the Higgs field acquires a non-zero VEV, namely $\langle h\rangle = h_0 \neq 0$. 
%The non-canonical interaction therefore can be naturally introduced by the kinetic term
%\begin{align}
%\vert \partial_\mu\Phi_H^\prime \vert^2 \rightarrow \vert \partial_\mu\Phi_H\vert^2  + \frac{\Phi_H^\dagger \Phi_H}{\Lambda^2} (\partial \phi)^2 + \cdots,
%\end{align}
%where $\Lambda$ now behaves like a symmetry-breaking scale for the additional $U(1)$ beyond the SM. 
{\footnote{
It is also desirable to specify the origin of the system \eqref{system} from explicit theories at high energy. For example,
the non-canonical kinetic interaction in \eqref{fieldspace1} can be found in the presence of an approximated $U(1)$ symmetry in the extended Higgs sector beyond the SM \cite{Gong:2012ri}.
This type of non-canonical kinetic interaction can also be found in inflation models based on supergravity with a soft-breaking of the shift-symmetry \cite{Yamaguchi:2011kg}.
We want to emphasize that the model \eqref{system} is used as an simple example for the heavy-lifting mechanism so that there is in fact no primary assumption for its origin.}
}

The cutoff scale of the system \eqref{system} is non-trivial since the target field space of $\phi$-$h$ can be curved.
For convenience, we perform the reparametrization for both fields as
\begin{align}
R = (\Lambda^2 +h^2)^{1/2}, \qquad\qquad \theta = \phi/\Lambda,
\end{align}
so that the kinetic part of the system becomes
\begin{align}
\mathcal{L} \supset -\frac{1}{2} R^2 \left(\partial_\mu \theta\right)^2 -\frac{1}{2}\frac{R^2}{R^2 - \Lambda^2} \left(\partial_\mu R\right)^2.
\end{align}
In this representation, the classical value of $R$ acts as the canonical radius for $\theta$, and the rescaled inflaton $\theta$ behaves as the angular mode in the polar coordinate system.
In general, the target field space is not flat since the radial mode $R$ is not canonically normalized. There are two interesting limits of this system.

\begin{enumerate}
	\item For $h^2 \ll \Lambda^2$, the radial mode $R \rightarrow \Lambda$ and the non-canonical $\phi$-$h$ interaction is suppressed by the factor $h^2/\Lambda^2 \ll 1$. In this limit the field space is nearly flat and it is nothing but the conventional single-field inflation with Higgs as an additional degree of freedom. We refer this regime as the flat-decoupling limit of the $\phi$-$h$ system (to be distinguished from the gravitational decoupling in the $\pi$-$\sigma$ model). 
	%This is the parameter space considered in \cite{Kumar:2017ecc}.
	
	\item For $h^2 \gg \Lambda^2$, the radial mode $R \rightarrow h$ coincide with the Higgs field. The field space is flat as the factor $R^2/(R^2 -\Lambda^2) \rightarrow 1$ becomes canonically normalized in the polar coordinate representation. 
	We refer this regime as the curvelinear limit of the $\phi$-$h$ system, since the inflationary trajectory is now curved.
	Heavy-lifting in this regime was not covered by previous literatures.
	%This is the limit similar to that of the quasi-single-field inflation with a constant turning trajectory.
\end{enumerate}

We now study the symmetry breaking of the Higgs potential in the system \eqref{system}.
If the Higgs field develops a well-defined background value $h_0$ during inflation, the homogeneous field equations of the system are given by
\begin{align} 
\label{eq:Friedmann_1}
3M_p^2H^2 &= \frac{1}{2} R_0^2 \dot{\theta}_0^2 +\frac{1}{2}\dot{h}_0^2 +V(\Lambda\,\theta_0) + \frac{\lambda}{4} h_0^4, \\
\label{eq:Friedmann_2}
-2M_p^2 \dot{H} &= R_0^2 \dot{\theta}_0^2 + \dot{h}_0^2, \\
\ddot{\theta}_0 +3 H\dot{\theta}_0 + \frac{2h_0\dot{h}_0}{R_0^2} \dot{\theta}_0+   \frac{V_\theta}{R_0^2}&= 0, \\
\ddot{h}_0 + 3H \dot{h}_0 + \lambda h_0^3 &=  h_0 \dot{\theta}_0^2, \label{eom:h}
\end{align}
where $R_0 = (\Lambda^2 +h_0^2)^{1/2}$ and $V_\theta \equiv \partial V/\partial \theta$.
For $\lambda > 0 $, there exists a static solution $h_0^2 = \dot{\theta}_0^2/\lambda$ such that the Higgs field 
%becomes classically decoupled from the inflationary dynamics once $h$ reaches equilibrium. 
can develops a non-zero vacuum expectation value (VEV) $h_0 = \pm \dot{\theta}_0/\sqrt{\lambda} $. 
This non-trivial VEV is sourced by the essential dynamics $\dot{\theta}^2 = \dot{\phi}^2/\Lambda^2$ of slow-roll inflation and is invariant under a constant shift of the inflaton value $\theta_0 \rightarrow \theta_0 + c$.
Expanding the effective potential
\begin{align}
V_{\rm eff} = \frac{\lambda}{4} h^4 -\frac{1}{2} \dot{\theta}_0^2 h^2,
\end{align}
around $h_0$ one finds the effective mass $m_h^2 = 2 \dot{\theta}_0^2$. In the gravitational decoupling limit the mass $m_h^2$ becomes a constant. A stable $h_0$ asked by the stochastic condition $m_h^2 \gtrsim H^2$ \cite{Starobinsky:1986,Starobinsky:1994bd} can be satisfied if $\dot{\theta}_0^2 \gtrsim H^2/2$.
Note that the first slow-roll parameter is related to the non-zero Higgs VEV as
\begin{align}
	\epsilon = - \frac{\dot{H}}{H^2} \simeq   \frac{R_0^2 \dot{\theta}^2}{2 M_p^2 H^2} = \frac{\lambda R_0^2 h_0^2}{2 M_p^2 H^2}.
\end{align}
In the flat-decoupling limit ($h_0^2 \ll \Lambda^2$) where $R_0 \approx \Lambda$, the slow-roll condition implies $\Lambda \ll 2 M_p H/\dot{\theta}_0$.
In the curvelinear limit ($h_0^2 \gg \Lambda^2$) where $R_0 \approx h_0$, the smallness of $\epsilon \simeq \lambda h_0^4/(2M_p^2 H^2)$ instead guarantees the subdominance of the Higgs vacuum energy to the background energy density.

\subsection{Scales of heavy Higgs}

The non-zero Higgs VEV $h_0$ is led by the slow-roll inflation dynamics $-2M_p^2 \dot{H} = R_0^2 \dot{\theta}_0^2$ and $3H \dot{\theta}_0 = -V_\theta/R_0^2$ at the first-order of $\epsilon$. We can treat $h_0$ as a stable constant during the slow-rolling of $\theta$ given that $\dot{h}_0$ is at least second-order in the slow-roll parameters.
%The non-zero Higgs VEV $h_0$ is very stable during inflation, given that $\dot{h}_0$ is at least second-order in the slow-roll parameters.
%This is consistent with the single-field inflation dynamics at the first-order of $\epsilon$ as $-2M_p^2 \dot{H} = R_0^2 \dot{\theta}_0^2$ and $3H \dot{\theta}_0 = -V_\theta/R_0^2$.
Performing the scalar perturbations $\theta(t, {\bf x}) = \theta_0(t) + \delta\theta(t, {\bf x})$ and $h(t, {\bf x}) = h_0(t) + \delta h(t, {\bf x})$ to \eqref{system}, we obtain the quadratic Lagrangian as
\begin{align}\label{quadratic perturbations}
\mathcal{L}_2 = \frac{1}{2} \left[R_0^2 \delta\dot{\theta}^2- \frac{R_0^2}{a^2}(\partial_i\delta\theta)^2 + \delta\dot{h}^2 -\frac{1}{a^2}(\partial_i \delta h)^2 - m_h^2 \delta h^2 + 4h_0 \dot{\theta}_0 \delta h \delta \dot{\theta}\right] + \mathcal{O}(\epsilon)\cdots,
\end{align}
where $\mathcal{O}(\epsilon)$ means quadratic perturbations that are suppressed by the slow-roll parameters (which includes the mass term of inflaton). 
The terms shown in \eqref{quadratic perturbations} can also be derived from the general perturbation theory, and one can check that metric perturbations only contribute to $\mathcal{O}(\epsilon)$.
%\footnote{Taking $\epsilon \rightarrow 0$ to obtain the part of Lagrangian decoupled from metric perturbations is sometimes called the decoupling limit with gravity \cite{Baumann:2011su}.} 

To see the dynamics of the system, it is convenient to use the canonically normalized field $\theta_c = R_0\theta$ with respect to the canonical commutation relation for canonical quantization.
The quadratic Lagrangian \eqref{quadratic perturbations} is rewritten as
\begin{align}\label{quadratic perturbations canonical}
\mathcal{L}_2 \supset \frac{1}{2} \left[ \delta\dot{\theta}_c^2- \frac{1}{a^2}(\partial_i\delta\theta_c)^2 + \delta\dot{h}^2 -\frac{1}{a^2}(\partial_i \delta h)^2 - m_h^2 \delta h^2 + 2\mu \delta h \delta \dot{\theta}_c\right],
\end{align}
where the mixing parameter 
\begin{align}
\mu \equiv \frac{2h_0 \dot{\theta}_c}{R^2} = \frac{2\dot{\theta}_0^2}{\sqrt{\dot{\theta}_0^2 + \lambda \Lambda^2}} ,
\end{align}
plays a key role in the dynamics of our system. Note that once $\lambda$ and $\Lambda$ are fixed, both $\mu$ and $m_h^2$ are controlled by the same parameter $\dot{\theta}$ so that they are not independent from each other. This is a fundamental difference from the constant-turn quasi-single-field inflation \cite{Chen:2009we,Chen:2009zp,An:2017hlx,Pi:2012gf} (or the strongly mixed $\pi$-$\sigma$ model \cite{Baumann:2011su}).   

We classify the $\phi$-$h$ system with respect to the mixing parameter $\mu$ as
\begin{itemize}
    \item weak-mixing: $\mu / H < 1$, and
    \item strong-mixing: $\mu / H > 1$.
\end{itemize}
For $H > \mu$, the interaction $2\mu \delta h \delta \dot{\theta}_c$ does not play an important role during inflation and the system \eqref{quadratic perturbations canonical} simply describes two weakly interacted scalar fields. In this case we expect the standard picture of a massless Higgs field.
The condition $\mu > H$ for the presence of a strong-mixing implies 
\begin{align}\label{cond:heavy-Higgs}
\dot{\theta}_0^2 > \frac{H^2}{8} + \frac{H^2}{8}\sqrt{1+16\frac{\lambda \Lambda^2}{H^2}}.
\end{align}

\begin{figure}
	\begin{center}
		\includegraphics[width=75mm]{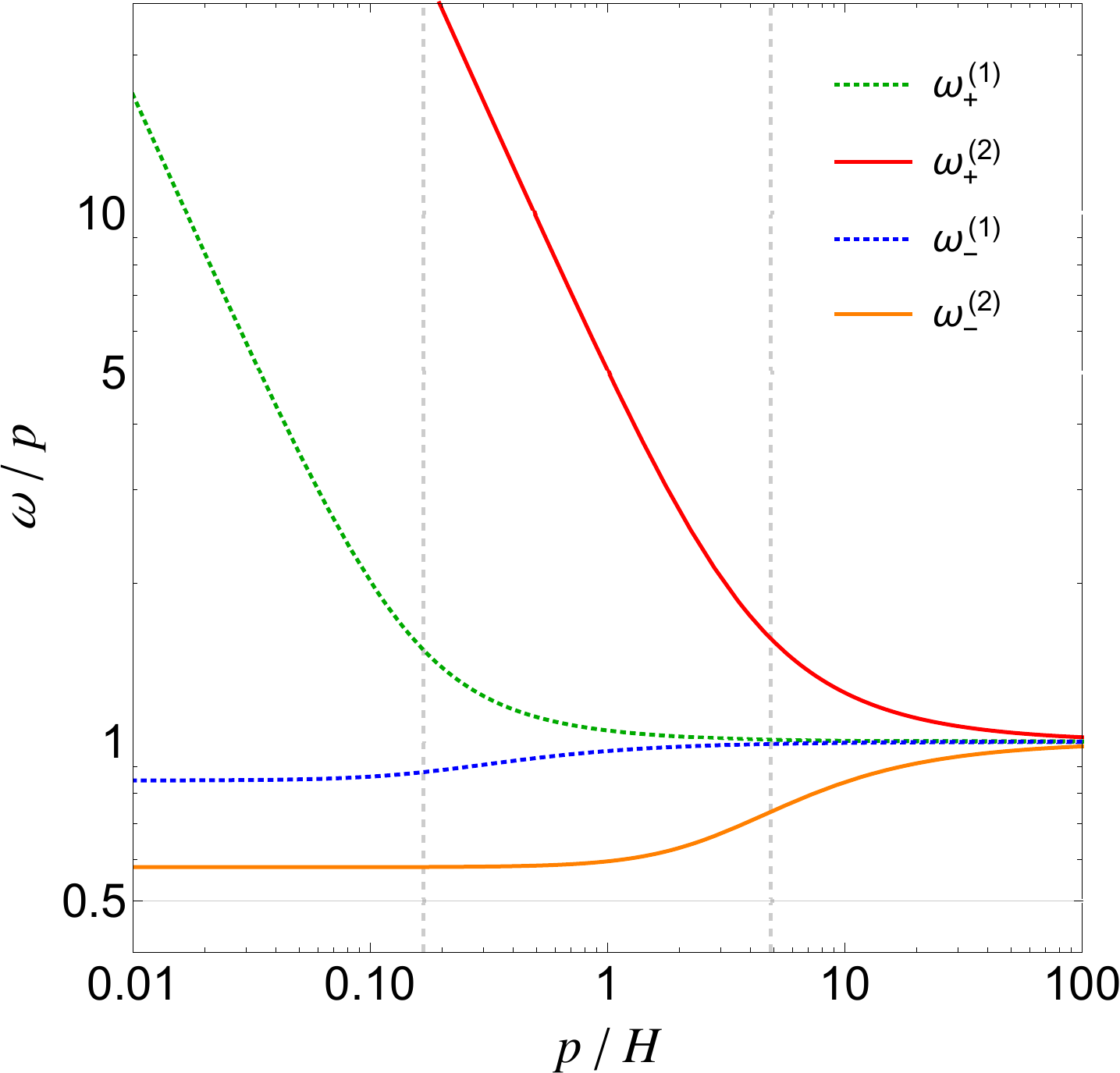}
		%\hfill
		%\includegraphics[width=75mm]{omega_p.eps}
	\end{center}
	\caption{\label{fig_omega_k}The change of the dispersion relation with respect to the physical wavenumber $p$ in the Hubble unit. For $\omega_\pm^{(1)}$ the parameters $\Lambda = 2H$, $\lambda =0.01$ and $\dot{\theta}_0 = 0.1 H$ are used, which gives $\mu^{(1)} \simeq 0.04 H $. For $\omega_\pm^{(2)}$ the parameters $\Lambda = 2H$, $\lambda =0.01$ and $\dot{\theta}_0 = 2 H$ are used, which gives $\mu^{(2)} \simeq 4 H $.
	}
\end{figure}

%For the $\phi$-$h$ system considered in this work $\mu$ and $m_h^2$ vanish together if one take $\dot{\theta} \rightarrow 0$.
To understand how the two degrees of freedom decoupled with the energy scales, let us write down the equations of motion of the perturbations
\begin{align} \label{eom: theta-h}
\delta\ddot{\theta}_c + 3H \delta\dot{\theta}_c+ \frac{k^2}{a^2} \delta\theta_c &= -\mu \left(\delta\dot{h} + 3H \delta h \right), \\
\delta\ddot{h} + 3H \delta\dot{h}+ \left(\frac{k^2}{a^2} + m_h^2 \right) \delta h &= \mu \delta\dot{\theta}_c.
\end{align}
In the long-wavelength regime with $k/a \rightarrow 0$, we expect the usual solution $\delta\theta_c \rightarrow$ constant and $\delta h \rightarrow 0$ of the single-field inflation. For $p = k/a \gg H$ we are allowed to neglect the cosmic expansion so that the equations of motion are reduced to
\begin{align}
\delta\ddot{\theta}_c + p^2 \delta\theta_c &= -\mu \delta\dot{h} , \\
\delta\ddot{h} +  \left(p^2 + m_h^2 \right) \delta h &= \mu \delta\dot{\theta}_c .
\end{align}
The solutions in the subhorizon regime thus take the form of $\delta\theta_c \sim \delta\theta_\pm e^{i \omega_\pm t}$ and $\delta h \sim \delta h_\pm e^{i \omega_\pm t} $ \cite{Achucarro:2012sm,Achucarro:2010jv}, where the two frequencies are found as 
\begin{align}\label{def:omega_pm}
\omega_\pm^2 &= p^2 + \frac{m_h^2+\mu^2}{2} \pm \sqrt{p^2\mu^2 + \frac{(m_h^2 + \mu^2)^2}{4}}, \\
 &= p^2 + \frac{m_h^2}{2 c_h^2} \pm \sqrt{p^2 \mu^2 + \frac{m_h^4}{4 c_h^4}}.
\end{align}
Here the speed of sound $c_h^2$ is simply defined in the limit of $p \ll (m_h^2 + \mu^2)/\mu$ such that the low-energy frequency can be expanded as
\begin{align}\label{eq:omega_neg}
\omega_-^2 \rightarrow c_h^2 p^2 + \mu^4 \frac{c_h^6}{m_h^6} p^4 = c_h^2 p^2 + (1- c_h^2)^2\frac{c_h^2}{m_h^2} p^4,
\end{align}
and thus it indicates
\begin{align}\label{def:cs2}
c_h^2 = \frac{m_h^2}{m_h^2 + \mu^2} = \frac{\dot{\theta}_0^2 + \lambda \Lambda^2}{3\dot{\theta}_0^2 + \lambda \Lambda^2}.
\end{align}
The result \eqref{def:cs2} appears to be the same as using the effective field approach for a curvelinear trajectory after neglecting the slow-roll parameter suppressed effective mass \cite{Achucarro:2012sm,Achucarro:2010jv}
With the definition \eqref{eq:omega_neg} the low-energy mode has a linear dispersion relation $\omega_- \approx c_h p$ for $p^2 \ll m_h^6 \mu^{-4} c_h^{-4}$ and a nonlinear dispersion relation $\omega_- \propto  p^2$ for  $p^2 \gg m_h^6 \mu^{-4} c_h^{-4}$.

The dispersion relation of the two frequency modes $\omega_\pm$ given by \eqref{def:omega_pm} is depicted in Fig. \ref{fig_omega_k}, yet keeping in mind that these solutions are only valid for subhorizon scales. The modes $\omega_\pm^{(1)}$ are in the case with $\mu < H$ and $\omega_\pm^{(2)}$ are in the case with $\mu > H$. For $\mu < H$, $\omega_\pm^{(1)}$ become almost degenerate at the Hubble scale during inflation ($p \sim H$) and they recover the usual linear dispersion relation $\omega \approx p$.

On the other hand, in the limit of $p \ll (m_h^2 + \mu^2)/\mu$ the high-energy mode
\begin{align}
\omega_+^{(2)} \rightarrow m_h/c_h,
\end{align}
describes a heavy degree of freedom during inflation as long as $m_h \gg c_h H$. Thus, with a mixing $\mu > H$, the existence of Higgs as a heavy field during inflation does not necessarily requires $m_h^2 \gtrsim H^2$. In fact, in the $\pi$-$\sigma$ model one can make $\sigma$ a heavy mode merely due to a strong-mixing $\mu/H \gg 1$ with a mass $m_\sigma \ll H $, provided that $c_h^2 \ll 1$ \cite{Baumann:2011su,Gwyn:2012mw}.
However, in our scenario the two parameters $\mu$ and $m_h$ are not independent, and one can check that $c_h^2 \rightarrow 1$ in the flat-decoupling limit of Higgs and inflaton where $\dot{\theta}_0^2 \ll \lambda \Lambda^2$ and $c_h^2 \rightarrow 1/3$ in the curvelinear limit where $\dot{\theta}_0^2 \gg \lambda \Lambda^2$. Based on these findings one can identify the energy scale to have a heavy Higgs field during inflation as
\begin{align}
\mu_h \equiv (m_h^2 + \mu^2 )^{1/2} = m_h/c_h,
\end{align}
where $\mu_h \rightarrow 0$ as $\dot{\theta}_0 \rightarrow 0$.
We therefore identify the \textit{heavy-Higgs condition} as: $\mu_h > H$, which is namely $m_h^2 > c_h^2 \geq H^2/9$. This implies the lower limit for a heavy Higgs as
\begin{align}
    \dot{\theta}_0 > H /\sqrt{18},
\end{align}
which is to be compared with \eqref{cond:heavy-Higgs}. 
%is self-consistent with the  condition ($\mu > H$) led by \eqref{cond:heavy-Higgs}.
The corresponding values of the examples $\mu_h^{(1)}/H \approx 0.17$ and $\mu_h^{(2)}/H \approx 4.88$ are given as the vertical lines in Fig. \ref{fig_omega_k}.

\subsection{Perturbativity and Naturalness}\label{subsec. naturalness}
\begin{figure}
	\begin{center}
		\includegraphics[width=50mm]{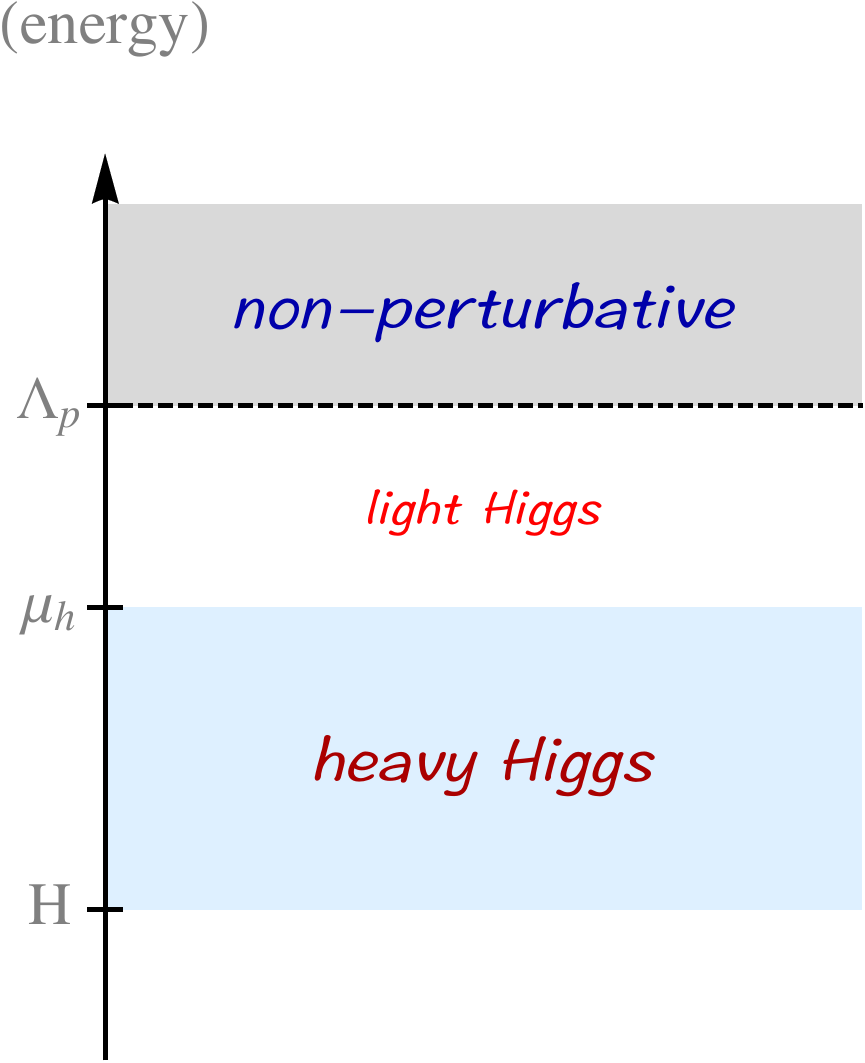}
		\qquad\qquad\qquad
		\includegraphics[width=50mm]{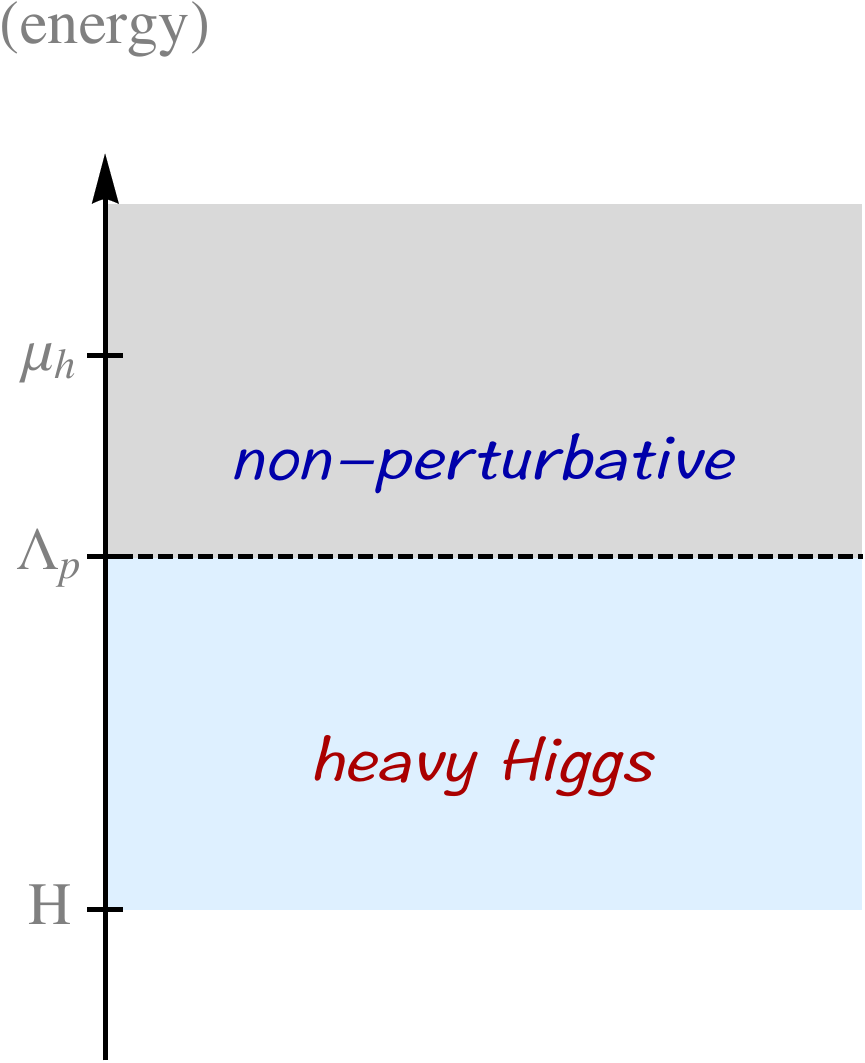}
	\end{center}
	\caption{\label{fig_energyscale}
		Illustration of the energy scales with two kinds of hierarchy.
	}
\end{figure}
%For $\mu^4 < R_0^2 \dot{\theta}_0^2 = - 2M_p^2 \dot{H}$, $\Lambda_t = (R_0\dot{\theta}_0)^{1/2}$.

%For $\mu^4 > R_0^2 \dot{\theta}_0^2 = - 2M_p^2 \dot{H}$, 
%\begin{align}
%\Lambda_t = \left(\frac{\mu}{m_h}c_s\right)^2 \frac{(R_0\dot{\theta}_0)^{4/7}}{(m_h/c_s)^{1/7}}.
%\end{align}

We now check if the perturbative expansion for higher-order terms are well-defined in the presence with a strong quadratic mixing $\mu > H$. We shall identify a scale $\Lambda_p$ as the cutoff for the higher-order expansion. If the condition $\Lambda_p > \mu_h$ can be satisfied, the Higgs field recovers the usual dispersion relation $\omega = k$ as a relativistic degree of freedom before the break down of the perturbative expansion of the theory. This case is illustrated by the left panel of Fig. \ref{fig_energyscale}.

Let us consider the cubic interactions introduced by the Higgs-inflaton coupling from \eqref{fieldspace1}
\begin{align}\label{cubic interactions}
\mathcal{L}_3 \supset \frac{h_0}{R_0^2} \left[\delta\dot{\theta}_c^2 -\frac{1}{a^2}(\partial_i \delta\theta_c)^2\right] \delta h
+ \frac{\dot{\theta}_c}{R_0^2}\delta h^2 \delta\dot{\theta}_c.  
\end{align}
With a linear dispersion relation $\omega = k$, the temporal derivative and spatial derivative has the same dimension so that we can easily identify $\Lambda_p = R_0^2/h_0$ from the first two cubic interactions in \eqref{cubic interactions}. Therefore $\Lambda_p > \mu_h$ implies that
\begin{align}
(\dot{\theta}_0^2 + \lambda \Lambda^2)^2 >  2 \lambda \dot{\theta}_0^4/c_h^2.
\end{align}
One can find that even in the curvelinear limit where $\dot{\theta}_0^2 \gg \lambda \Lambda^2$ the condition holds if $\lambda < 1/6$. 
%or in the decoupling limit where $\dot{\theta}_0^2 \ll \lambda \Lambda^2$ this condition holds if $\Lambda > (2/\lambda)^{1/3} \dot{\theta}_0$.
The coefficient of the third interaction in \eqref{cubic interactions} is dimensionless so that perturbativity simply requires $\dot{\theta}_c/R_0^2 < 1$,
which asks %$\Lambda > \dot{\theta}_0$ in the decoupling limit or 
$\lambda < 1$ in the curvelinear limit. 

Note that in the case with $\mu_h > \Lambda_p $, the system may enter to the non-perturbative region with a non-linear dispersion given by \eqref{eq:omega_neg}, as illustrated by the right panel of Fig. \ref{fig_energyscale}. In this case the space and time coordinate can have different dimensions, making the discussion much complicated. To determine the cutoff scale $\Lambda_p$, we write down the non-relativistic version of the Lagrangian \eqref{quadratic perturbations canonical} as
\begin{align}\label{quadratic perturbations non-relativistic}
\mathcal{L}_2 \approx
\mu \delta h \delta \dot{\theta}_c - \frac{1}{2 a^2} (\partial_i \delta\theta_c)^2 -\frac{1}{2a^2} (\partial_i \delta h)^2 -\frac{1}{2}m_h^2 \delta h^2,
\end{align}
where time-derivatives have dropped out. 

For $p \gg m_h/(1-c_h^2)$, the low-energy mode $\omega_- \approx (1-c_h^2)(c_h/m_h) p^2 \propto p^2/m_h$ so that the energy density has the dimension $[T^{00}] = [\omega_-][k]^3 = [\omega_-]^{5/2}[m_h]^{3/2}$. Identifying each term in \eqref{quadratic perturbations non-relativistic}, we find that $[m_h] = [\delta h]= [\delta\theta_c] =[\omega_-]$. Rescaling the spatial coordinate as $x_r = \mu^{1/2} x$ with the redefinition of fields as $\delta h_r = \mu^{-1/4} \delta h$ and $\delta \theta_r = \mu^{-1/4} \delta\theta_c$, the momentum conjugate is normalized as 
$p_{\delta \theta_r} \equiv \partial \mathcal{L}_2/ \partial \delta \dot{\theta}_r = \delta h_r $.
The cubic interactions are thus parametrized as
\begin{align}
\mathcal{L}_3 \supset \frac{1}{\left(\Lambda_{p1}\right)^{7/4}} \delta \dot{\theta}_r ^2 \delta h_r 
- \frac{1}{\left(\Lambda_{p2}\right)^{3/4}} \left(\tilde{\partial}_i \delta \theta_r\right)^2 \delta h_r 
+\frac{1}{\left(\Lambda_{p3}\right)^{3/4}} \delta\dot{\theta}_r \delta h_r^2,
\end{align}
where $\tilde{\partial_i}$ means the spatial derivative with respect to $x_r$, and 
\begin{align}
\Lambda_{p1} = h_0^{4/7} \mu^{3/7}, \quad \Lambda_{p2} =  h_0^{4/3} \mu^{-1/3} \quad \text{and} \quad \Lambda_{p3} = \mu \lambda^{-2/3}.
\end{align}
Here the curvelinear condition $R_0 \rightarrow h_0$ is used. Since each loop integral of the cubic interactions asks at least two insertions of the vortex, the cutoff is raised by a factor of $16\pi^2$ \cite{Baumann:2011su}. As a result, one can check that in order to realize $\mu_h > \Lambda_p$, the cutoff scales $\Lambda_{p1}$, $\Lambda_{p2}$ and $\Lambda_{p3}$ all ask $\lambda \gg 1$. These results are inconsistent with the parameter space of our consideration. In summary, for $\lambda \ll 1$ the two fields in the system always become weakly coupled before they reach the non-perturbative region. 

\begin{figure}
	\begin{center}
		\includegraphics[width=100mm]{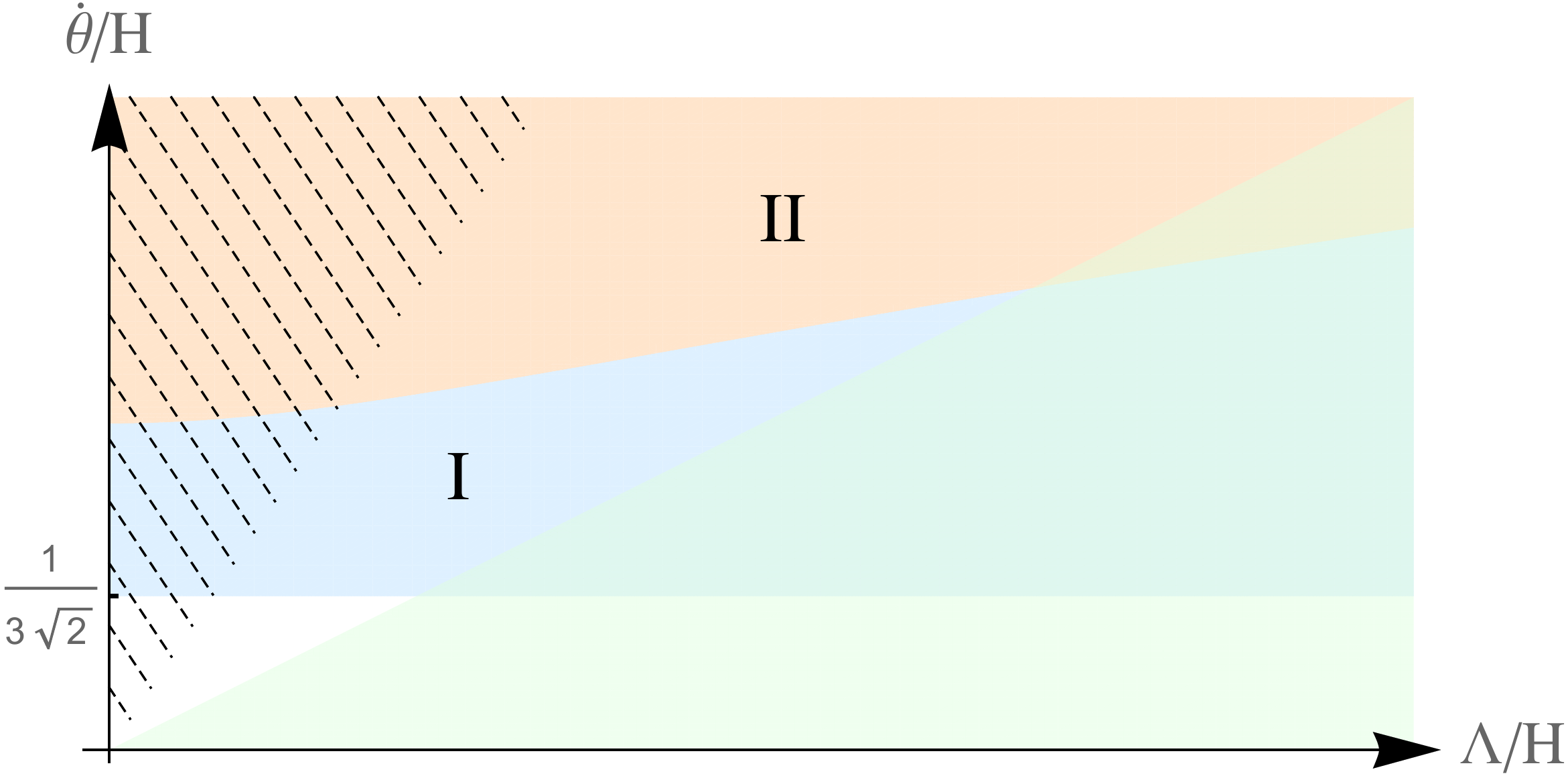}
	\end{center}
	\caption{\label{fig_naturalness}
		Parameter space for the $\phi$-$h$ system with $\lambda = 0.01$. The green area is the flat-decoupling limit with $\dot{\theta}/\Lambda < \sqrt{\lambda}$. The meshed area is incompatible with the Naturalness condition $\dot{\theta}/\Lambda < 1$. The blue (orange) area is the region satisfies the heavy Higgs condition $\dot{\theta}/H > \sqrt{18}$ with a weak-mixing (strong-mixing) $\mu/H <1$ ($\mu/H > 1$), respectively. 
	}
\end{figure}

One can impose the condition $\dot{\phi}/\Lambda^2 < 1$ to suppress the higher-order corrections from $(\partial_\mu \phi)^n/\Lambda^n$ to the system \eqref{system}. In terms of $\theta = \phi/\Lambda$, the Naturalness condition, namely $\dot{\theta} < \Lambda$, includes all the flat-decoupling region $\dot{\theta} < \sqrt{\lambda}\Lambda$ since the perturbativity asks $\lambda < 1$. As shown in Fig. \ref{fig_naturalness}, a heavy Higgs field can be realized in the curvelinear limit with respect to the perturbativity and Naturalness in Region I with a weak quadratic mixing, or in Region II with a strong quadratic mixing.

\section{Observational impact of heavy Higgs}\label{Sec. imprints}
\subsection{power spectrum}

\noindent
\textbf{Observational constraints.}
We study in this section the corrections to the power spectrum led by the Higgs-inflaton interactions at linear order. Given that the quadratic interaction of the system \eqref{quadratic perturbations canonical} is a derivative coupling, these corrections are suppressed on superhorizon scales so that they do not change the scale dependence of the power spectrum. However, in the strong-mixing regime these corrections may become comparable to the leading term \cite{An:2017hlx} and thus one has to rescale the amplitude to meet the observational constraint. Treating the background value $\dot{\theta}_0$ as a free parameter, we shall rescale the power spectrum with respect to the observational constraint 
\begin{align}
    \Delta_\zeta^2 \equiv \frac{H_\ast^2}{ R_\ast^2\,\dot{\theta}_\ast^2}P^\ast_{\theta} \approx 2.2 \times 10^{-9},
\end{align}
where $P^\ast_{\theta} = H_\ast^2/(4\pi^2)$ is a reference field spectrum for a given reference angular velocity $\dot{\theta}_\ast$, which is more convenient to choose in the flat-decoupling limit (namely $\dot{\theta}_\ast \ll \sqrt{\lambda}\Lambda$) such that $R_\ast = \Lambda$. $H_\ast$ is the reference Hubble parameter as viewed in the flat-decoupling limit. To keep a constant $\Delta_\zeta^2$ while varying $\dot{\theta}_\ast$, we shall adjust parameters in $P^\ast_{\theta}$. 
In general, we can rescale $H_\ast$ or $\Lambda$.

\begin{figure}
	\begin{center}
		\includegraphics[width=100mm]{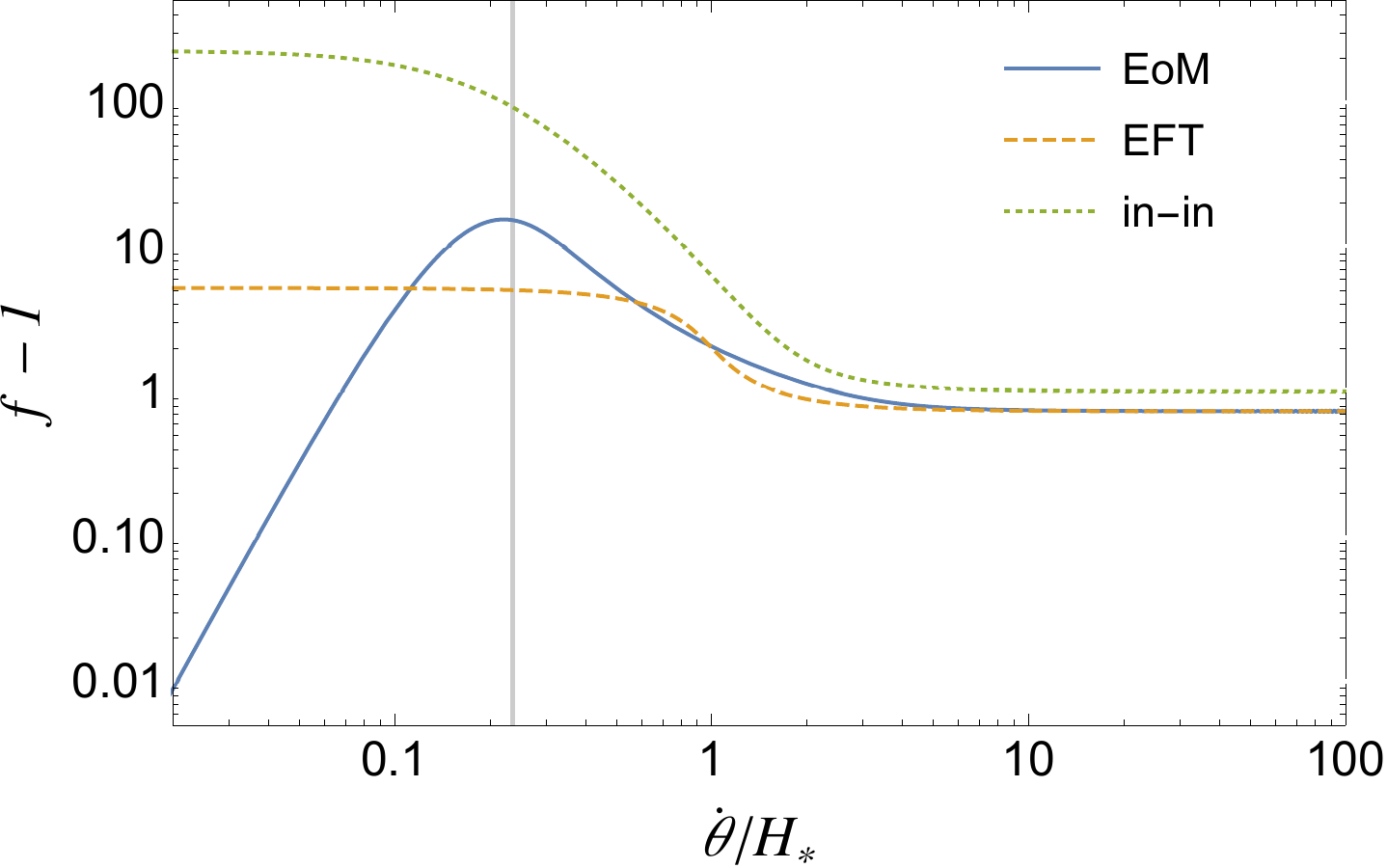}
	\end{center}
	\caption{\label{fig_powerspectrum}
		%The ratio of the Higgs correction to power spectrum $\Delta P_\zeta$ to the standard prediction given by single-field inflation $P^\ast_\zeta$ in terms of the coupling parameter $\dot{\theta}_0/H_\ast$. The vertical line describes the threshold value for the heavy Higgs condition.
		The ratio $f= P_\theta/P^\ast_\theta$ of the power spectrum with respect to the parameter $\dot{\theta}_0/H_\ast$ in the $H$-rescaling scheme. The vertical line describes the threshold value for the heavy Higgs condition.
	}
\end{figure}

\begin{figure}
	\begin{center}
		\includegraphics[width=100mm]{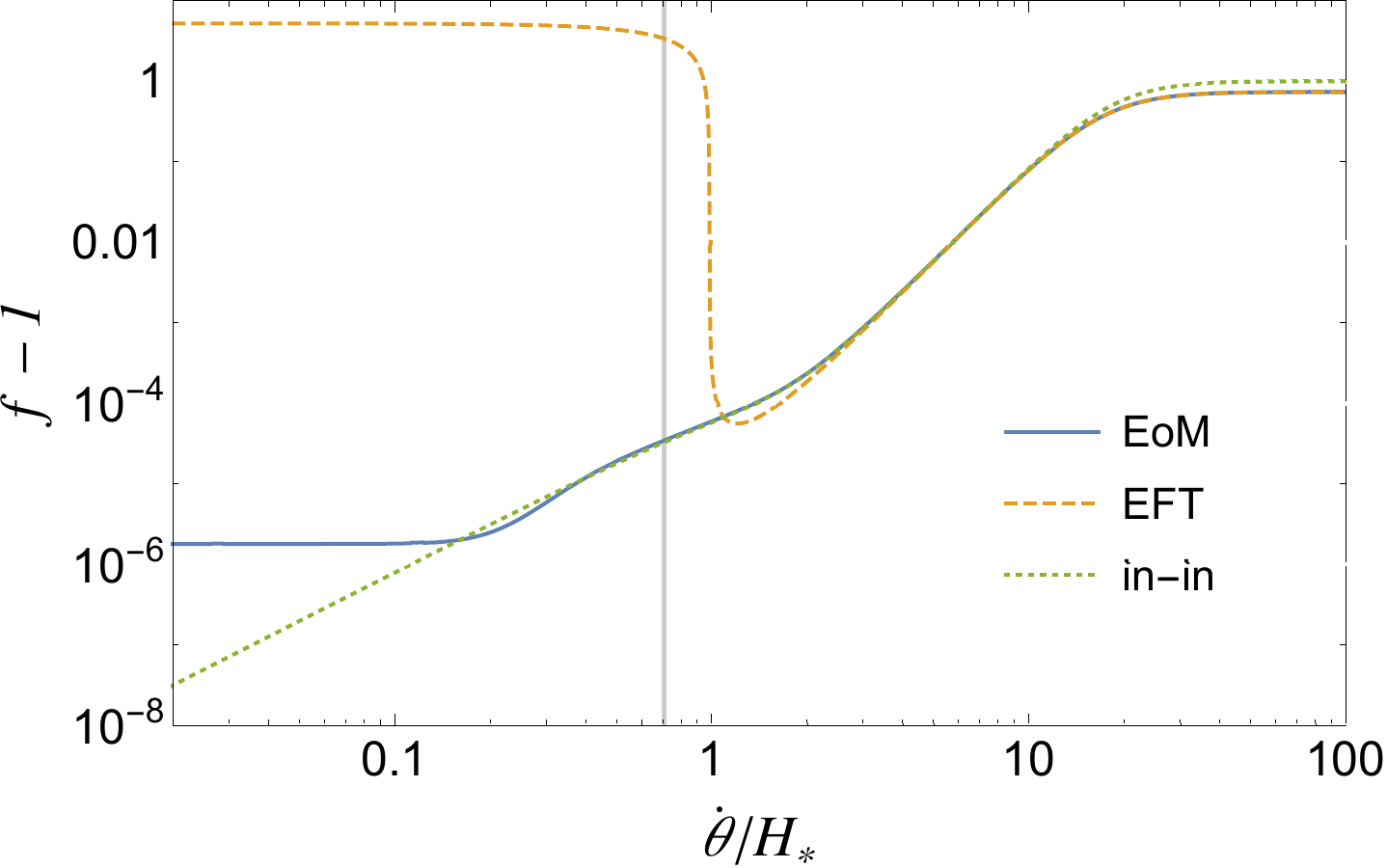}
	\end{center}
	\caption{\label{fig_powerspectrumL}
		The ratio $f= P_\theta/P^\ast_\theta$ of the power spectrum with respect to the parameter $\dot{\theta}_0/H_\ast$ in the $\Lambda$-rescaling scheme. The vertical line describes the threshold value for the heavy Higgs condition.
	}
\end{figure}

\begin{itemize}
\item
\textit{$H$-rescaling.}
The first choice is to rescale $H_\ast$ for a given $\dot{\theta}_0$. 
For numerical convenience, the change of the field spectrum evaluated under a constant Hubble parameter $H_\ast$ is parametrized as $P_\theta = f P_{\theta}^\ast$. We define a rescaled Hubble parameter $H$ such that
\begin{align}\label{H_rescaling}
    \Delta_\zeta^2 = \frac{H^2}{R^2_0\,\dot{\theta}_0^2}P_\theta^{\rm\, new} 
    = \frac{H^2}{R_0^2\,\dot{\theta}_0^2} \frac{H^2}{H_\ast^2} f  P_{\theta}^\ast,
\end{align}
where the rescaled field spectrum is $P_\theta^{\rm new} \equiv P_\theta H^2/H^2_\ast = H^2f/(4\pi^2)$. The rescaled Hubble parameter is then given by
\begin{align}
    \frac{H}{H_\ast} = 
    \left(\frac{\dot{\theta}_0}{\dot{\theta}_\ast}\right)^{1/2}
    \left(\frac{R_0}{R_\ast}\right)^{1/2} f^{-1/4},
\end{align}
%Note that one can redefine $P^{\rm new}_\theta \equiv H_{\rm new}^2 f P^\ast_\theta /H^2$ to justify the physical scale $H$. Once $\Lambda$ is fixed, we have $H_\ast = 2\pi \Lambda \Delta_\zeta \dot{\theta}_0/H$. 
The numerical results for $f$ with $\Lambda = 2H$ and $\lambda = 0.01$ are given in Fig. \ref{fig_powerspectrum}, where the threshold value for the heavy Higgs condition is $\dot{\theta}_0/H_\ast = 0.24$ and $\dot{\theta}_0/H_\ast > 0.27$ becomes strong-mixing.
%\footnote{For numerical convenience we in fact keep the input value $H_\ast$ unchanged and rescale $H$. In this sense, the physical interpretation of $H_\ast$ exchanges with $H$.}
We refer this method as the $H$-rescaling scheme.

\item
\textit{$\Lambda$-rescaling.}
Another choice to fit the observational constraint is to rescale $\Lambda$ with the change of $\dot{\theta}_0$. For the flat-decoupling case, we can define that
\begin{align}\label{L_rescaling}
    \Delta_\zeta^2 = \frac{H_\ast^2}{\Lambda^2 \,\dot{\theta}_0^2} P_\theta^{\rm \, new}
    = \frac{H^2_\ast}{\Lambda^2 \,\dot{\theta}_0^2}\frac{H^2}{H_\ast^2}P_\theta 
    %= \frac{H^2_\ast}{\Lambda^2 \,\dot{\theta}_0^2}P_\theta 
    = \frac{H^2_\ast}{\Lambda^2 \,\dot{\theta}_0^2} \frac{H^2}{H_\ast^2} f  P_\theta^\ast,
\end{align}
where the possible change of the power spectrum is absorbed in $H^2 = H_\ast^2 f^{-1}$ for convenience.
In this scheme $\Lambda$ is determined by a given $\dot{\theta}_0$ as $\Lambda = H_\ast^2/(2\pi\Delta_\zeta \dot{\theta}_0)$. We refer this method as the $\Lambda$-rescaling scheme, and the numerical results for $f$ with $\lambda = 0.01$ are given in Fig. \ref{fig_powerspectrumL}. For $\lambda = 0.01$, the threshold value for the heavy Higgs condition $\mu_h/H_\ast = 1$ gives $\dot{\theta}_0/H_\ast = 0.7$, the Naturalness condition asks $\dot{\theta}_0/H_\ast < 58.3$ and the strong-mixing $\mu/H_\ast > 1$ requires $\dot{\theta}_0/H_\ast > 5.5$. 
\end{itemize}


For both rescaling schemes the results from three kinds of approaches are summarized in Figures \ref{fig_powerspectrum} and \ref{fig_powerspectrumL}.
We discuss the implication from each approach as the follows.

\smallskip
\noindent
\textbf{Equation of Motion (EoM).}
The equation of motion (EoM) approach \cite{Chen:2015dga} solves quantum field fluctuations from a complete set of initial states that satisfy the canonical commutation relation. The Bunch-Davies vacuum states are special examples of these initial states and are usually applied to define as the vacuum of ``free fields'' for the in-in formalism \cite{Chen:2010xka,Wang:2013eqj,Weinberg:2005vy} in the interaction picture. However, according to the first-principles of the in-in formalism, these initial states in general need not to be fully decoupled from each other, and therefore the EoM approach is also useful to deal with mixed initial states arised from a strongly-coupled system. Initial mode functions for the $\theta$-$h$ system \eqref{quadratic perturbations canonical} are found to be
\begin{align}\label{mode EoM}
\delta \theta^\pm_k = \frac{H_\ast}{ \sqrt{4k^3}}e^{-i k \eta}(-k \eta)^{1\pm i \mu/(2H_\ast)}, \quad \text{and}\quad \delta h^\pm_k = \pm i \delta \theta^\pm_k,
\end{align}
where the derivation is given in Appendix \ref{Appx_EoM}. The dimensionless spectrum is given by $P_\theta = k^3(\vert\delta\theta_k^+\vert^2+\vert\delta\theta_k^-\vert^2)/(4\pi^2)$.

For both rescaling scheme, in the limit $\dot{\theta}_0/H_\ast \ll 1$ Higgs behaves as a light isocurvature mode with negligible corrections to the power spectrum. For $\dot{\theta}_0/H_\ast \gg 1$ the EoM results agree with the prediction from the effective field theory (EFT) method by integrating out the heavy Higgs field. In the intermediate regime $\dot{\theta}_0/H_\ast \sim \mathcal{O}(1)$, the numerical result from EoM method in the $H$-rescaling scheme cannot be explained by the analytical approach (in-in formalism) nor the EFT method.
We remark that the numerical results shown in Figs. \ref{fig_powerspectrum} and \ref{fig_powerspectrumL} do not diverge in the limit of $m_h \rightarrow 0$ where $\dot{\theta}_0/H_\ast \rightarrow 0$, since the parameter $\mu$ vanishes identically in this limit.

\smallskip
\noindent
\textbf{In-in formalism (in-in).}
As a crosscheck of our numerical results, we provide the analytical computation for the power spectrum based on the in-in formalism. The formulae for the $\phi$-$h$ system \eqref{quadratic perturbations canonical} computation can be found in \cite{Chen:2009zp,Chen:2012ge,Saito:2018omt}, and for inflaton with a generalized speed of sound $c_\theta^2 \leq 1$ is considered in \cite{Iyer:2017qzw,Lee:2016vti}. As given by the dotted line in Fig. \ref{fig_powerspectrum}, the correction from the quadratic interaction $\delta\mathcal{L}_2 = \mu \delta h \delta\dot{\theta}_c$ up to first order in the perturbative expansion is led by the form
\begin{align}\label{result: in-in}
\frac{\Delta P_\zeta}{P_\zeta^\ast} = 2\left(\frac{\mu}{H_\ast}\right)^2 
\left\lbrace \frac{\pi^2}{4 \cosh  \pi L} 
+ \frac{e^{\pi L}}{16\sinh \pi L} \,\Re \left[\psi^{(1)}\left(\frac{3}{4}+i \frac{L}{2}\right)-\psi^{(1)}\left(\frac{1}{4}+i \frac{L}{2}\right)\right] \right.\nonumber\\
\left. - \frac{e^{-\pi L}}{16\sinh \pi L} \,\Re \left[\psi^{(1)}\left(\frac{3}{4}- i \frac{L}{2}\right)-\psi^{(1)}\left(\frac{1}{4}- i \frac{L}{2}\right)\right]
\right\rbrace,
\end{align}
where $\psi^{(1)}(x) \equiv d^2 \ln \Gamma(x)/dx^2$ is the polygamma function, and
\begin{align}\label{def:L}
L = \sqrt{\frac{m_h^2}{H_\ast^2}-\frac{9}{4}}.
\end{align}
$\Delta P_\zeta = P_\zeta - P_\zeta^{\ast}$ is the deviation of the power spectrum from the expectation value of the standard single-field inflation $P_\zeta^\ast = \Delta_\zeta^2$.
%$P_\zeta^{\ast} = H^4/(4\pi^2 \Lambda^2 \dot{\theta}^2_0)$.
Therefore we have $f = \Delta P_\zeta/P_\zeta^\ast +1$.
One can check that $\Delta P_\zeta/ P_\zeta^\ast \rightarrow 1$ in the limit of $\dot{\theta}_0/H_\ast \rightarrow \infty$ where $m_h^2\rightarrow \infty$. The deviation from the EoM results in the strong-mixing regime is due to the fact that $\delta\theta$ aquires an effective speed of sound $c_\theta^2 < 1$ (see the discussion in EFT method).
Note that the expression \eqref{result: in-in} does not diverge in the limit of $\dot{\theta}_0/H_\ast \rightarrow 0$ since $\mu$ also goes to zero. This is one of the essential difference from the quasi-single-field inflation results \cite{An:2017hlx,Chen:2009we,Chen:2009zp}. 

\smallskip
\noindent
\textbf{Effective field theory (EFT).}
Given that Higgs behaves as a heavy degree of freedom in the strong-mixing limit ($\mu > H$), one may consider to integrate out the heavy Higgs to obtain an effective single-field theory. We denote the integrate-out process of the heavy Higgs as the effective field theory (EFT) approach \cite{Gong:2013sma,Achucarro:2010jv,Achucarro:2012sm,An:2017hlx,Gwyn:2012mw,Iyer:2017qzw,Tong:2017iat}, which is to be distinguished from the effective theory discussed in Section \ref{Sec. EFT}. 

The EFT approach for the two-field system \eqref{quadratic perturbations canonical} has been studied in the large mass regime ($m_h \gg \mu$ with $m_h \gg H$) \cite{Achucarro:2010jv,Achucarro:2012sm} and in the large mixing regime ($m_h \ll \mu$ with $\mu \gg H$) \cite{Baumann:2011su,An:2017hlx}. In our case, both $m_h$ and $\mu$ are controlled by the parameter $\dot{\theta}_0$ so that the generalized method \cite{Iyer:2017qzw,Gwyn:2012mw,Tong:2017iat} applied for $(m_h^2 + \mu^2)^{1/2} > H$ is required. Note that the effective mass for the heavy mode $\omega_+$ is nothing but $(m_h^2 + \mu^2)^{1/2} $ when taking $k \rightarrow 0$ into \eqref{def:omega_pm}. To derive the effective action with a modified speed of sound, one solves the heavy Higgs field in the conformally flat space with $a = -1/(H\eta)$ \cite{Tong:2017iat}. The solution is
\begin{align}\label{delta h_EFT}
\delta h = -\frac{a \mu R_0}{\partial_i^2 - a^2 (m_h^2 - 2H^2)} \delta \theta^\prime + \cdots,
\end{align}
where a prime is the derivative with respect to the conformal time $\eta$. 

The term shown in \eqref{delta h_EFT} only gives the local effect, while non-local effects must come from higher-order corrections. Using \eqref{delta h_EFT} we find the effective action
\begin{align}
S_{\rm eff} = \int d^4x \,\frac{a^2}{2} \left[\frac{R_0^2 }{c_{\theta}^2} \delta \theta^{\prime 2} -R_0^2 (\partial_i \delta \theta)^2\right],
\end{align}
where the effective speed of sound after integrating out the heavy Higgs reads
\begin{align}
c_{\theta}^{-2}(k\eta) = 1+ \frac{\mu^2}{H^2k^2\eta_\ast^2 +m_h^2 -2H^2},
\end{align}
where $\eta_\ast = - \beta/(k c_{\theta})$ is taken to be the value at the sound horizon crossing. The value of $\beta$ can be determined by matching the analytical solutions obtained in the $m_h \rightarrow 0$ limit, which gives $\beta^{1/2} = \Gamma(-1/4)^2/(16\pi)$ \cite{Gwyn:2012mw,An:2017hlx,Tong:2017iat}.
In the strong-mixing limit with $\dot{\theta}_0/H \gg 1$, the speed of sound reduces to the prediction \eqref{def:cs2} as
$c_{\theta}^{-2} \rightarrow c_h^{-2} = 1 + \mu^2/m_h^2 \approx 3$. In the weak-mixing limit with $\dot{\theta}_0/H \ll 1$, the speed of sound approaches to a constant value $c_{\theta}^{-2} \rightarrow 2/ \beta^2$. The power spectrum is enhanced by $P_\zeta = P_\zeta^\ast/c_{\theta} $ where $c_{\theta}^{-1} \approx 6.19$.

\subsection{bispectrum}

\begin{figure}
	\begin{center}
		\includegraphics[width=150mm]{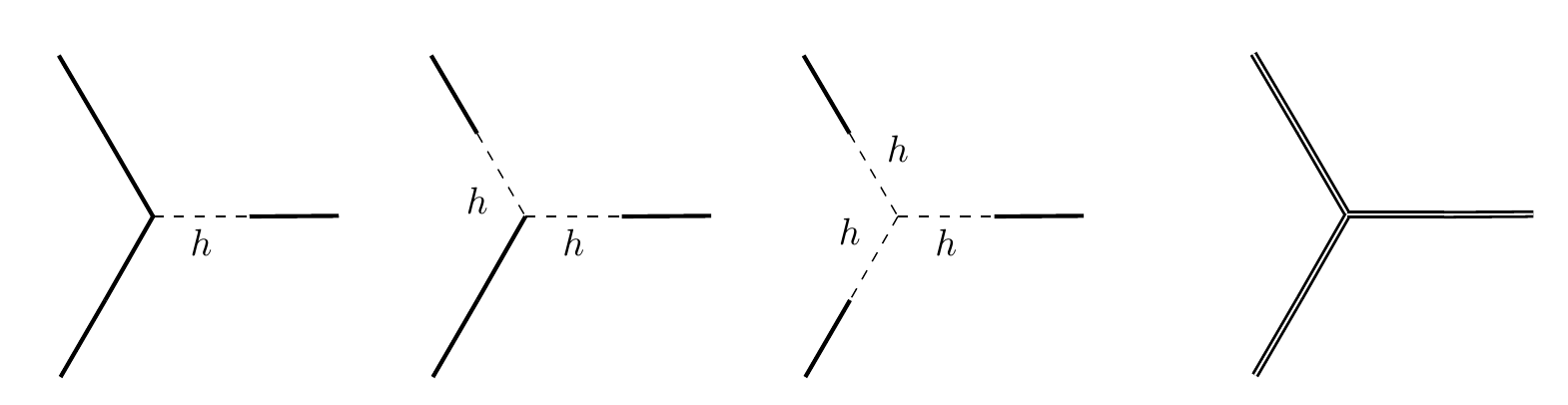}
	\end{center}
	\caption{\label{fig_Feynmen diagrams} Diagrams in the weak-mixing limit for the single (left), double (middle-left) and triple (middle-right) $\phi$-$h$ exchange processes in the three-point correlation functions. In the strong-mixing limit, the triple exchange diagrams converge to the contact diagram given by the right panel, where double solid lines denote the mixed $\phi$-$h$ propagator \eqref{eq:2-point mix}.}
\end{figure}

We now investigate the non-Gaussian correlation functions sourced by the Higgs field. Based on the results of power spectrum, we find that the numerical computation from the $\Lambda$-rescaling scheme is highly consistent with the prediction of in-in formalism for the intermediate region $\dot{\theta}_0/H\sim \mathcal{O}(1)$ where the EFT approach is not yet ready to apply. On the other hand, the numerical results from the $H$-rescaling scheme for the intermediate region can have distinctive behavior from the analytic approaches. The heavy Higgs in this range mainly corresponding to the curvelinear limit with a strong-mixing as labeled by the Region II in Fig. \ref{fig_naturalness}. We therefore focus on the study in this region by using the numerical method. As the simplest example, we compute the three-point correlation functions involved with the $\phi$-$h$ exchange processes led by \eqref{cubic interactions}.
In general, one can classify the cubic interactions of \eqref{cubic interactions} into
\begin{align}
\mathcal{L}^{(1)}_3 &= -a^3 \frac{h_0}{R_0^2}\left[\delta\dot{\theta}_c^2 -\frac{1}{a^2}(\partial_i \delta\theta_c)^2\right] \delta h, \\
\mathcal{L}^{(2)}_3 &= -a^3 \frac{\dot{\theta}_0}{R_0} \delta\dot{\theta}_c \delta h^2, \\
\mathcal{L}^{(3)}_3 &= -a^3 \lambda h_0 \delta h^3,
\end{align}
where $\mathcal{L}^{(1)}_3$,  $\mathcal{L}^{(2)}_3$ and $\mathcal{L}^{(3)}_3$ correspond to the vertices of the single, double and triple exchange diagrams, as depicted in Fig. \ref{fig_Feynmen diagrams}.

\smallskip
\noindent
\textbf{Equilateral limit.}
For $\mu < H$ the quadratic interaction $\delta \mathcal{L}_2 = \mu \delta h \delta\dot{\theta}_c$ is treated perturbatively. We can estimate the size of the non-Gaussianity in the equilateral limit by using the usual in-in formalism \cite{Chen:2009zp}. For the triple exchange diagram (the middle-right panel of Fig. \ref{fig_Feynmen diagrams}), the non-Gaussianity is estimated by
\begin{align}
   f_{NL} \sim \frac{1}{\Delta_\zeta} \left(\frac{\mu}{H}\right)^3 \frac{\lambda h_0}{H}.
\end{align}
In the curvelinear region $\mu \approx 2\dot{\theta}_0$ and $\Delta_\zeta \simeq H^2/(2\pi h_0\dot{\theta}_0)$ as $R_0 \approx h_0$, we find that
\begin{align}
    f_{NL} \sim 16\pi \left(\frac{\dot{\theta}_0}{H}\right)^6.
\end{align}
We expect this $f_{NL} \ll 1$ since it is the estimation for $\mu/H <1$, and thus $\dot{\theta}_0 < H/2$. One can perform a similar estimation to obtain that $f_{NL} \sim 4\pi (\dot{\theta}_0/H)^4$ for the double exchange diagrams and $f_{NL} \sim \pi (\dot{\theta}_0/H)^2$ for the single exchange diagrams.

For the strong-mixing case ($\mu > H$) the quadratic interaction $\delta \mathcal{L}_2$ also plays an important role in the equation of motion so that $\delta \theta$ and $\delta h$ are not treated independently.
We have shown in the previous section that the perturbativity of all cubic $\phi$-$h$ interactions in \eqref{cubic interactions} are well-defined even if the quadratic perturbations appear to be strongly coupled. Therefore our computation for the bispectrum can be taken as a modified in-in formalism based on the EoM approach given by Appedix \ref{Appx_EoM}. To be more precisely, we take solutions of the coupled EoM from quadratic perturbations to be the mode functions in the interaction picture and we treat cubic interactions as perturbations, following the scheme of \eqref{EoM_dec_modified}.
%\begin{align}
%\tilde{H}[\delta\theta, \delta h, p_\theta, p_h; t] = \tilde{H}_0[\delta\theta, \delta h, p_\theta, p_h; t] + \tilde{H}_I[\delta\theta, \delta h, p_\theta, p_h; t], \nonumber
%\end{align}

The three-point diagrams given by Fig. \ref{fig_Feynmen diagrams} are computed as
\begin{align}\label{eq:3point-commutator}
\nonumber
\langle \delta\theta^3(t)\rangle 
%&= \left\langle\left( \sum_{N=0}^{\infty}i^N \int^z dz_N \cdots \int^{z_2}dz_1 
%[H_{I,3}(z_1), \cdots [H_{I,3}(z_{N}),\delta\theta^3_I(z)]] \right) \right\rangle \\
%&
= i \int^{t}_{t_0}dt_1 \left\langle 0\vert  \left[\tilde{H}_{I,3}(t_1), \delta\theta^3_I(t)\right] \vert 0\right\rangle ,
\end{align}
where $\tilde{H}_{I,3}$ collects all cubic interactions \eqref{cubic_interactions_Hami} with $\delta\theta_I$ and $\delta h_I$ resolved from the EoM approach.
We adopt the conventional definition for the bispectrum $B_\theta$ as
\begin{align}
\langle \delta\theta_{\bf k_1}(t)\delta\theta_{\bf k_2}(t)\delta\theta_{\bf k_3}(t) \rangle 
&\equiv (2\pi )^3 \delta^3({\bf k_1+ k_2 + k_3}) B_\theta({ k_1, k_2 ,k_3}), \\
&=  (2\pi )^7 \delta^3({\bf k_1+ k_2 + k_3}) \frac{P_\theta^2}{(k_1k_2k_3)^2} S_\theta({ k_1, k_2 ,k_3}),
\end{align}
where $S_\theta$ is the dimensionless shape function and $P_\theta = P_\zeta\times(\dot{\theta}_0^2/H^2)$.

\begin{figure}
	\begin{center}
		\includegraphics[width=100mm]{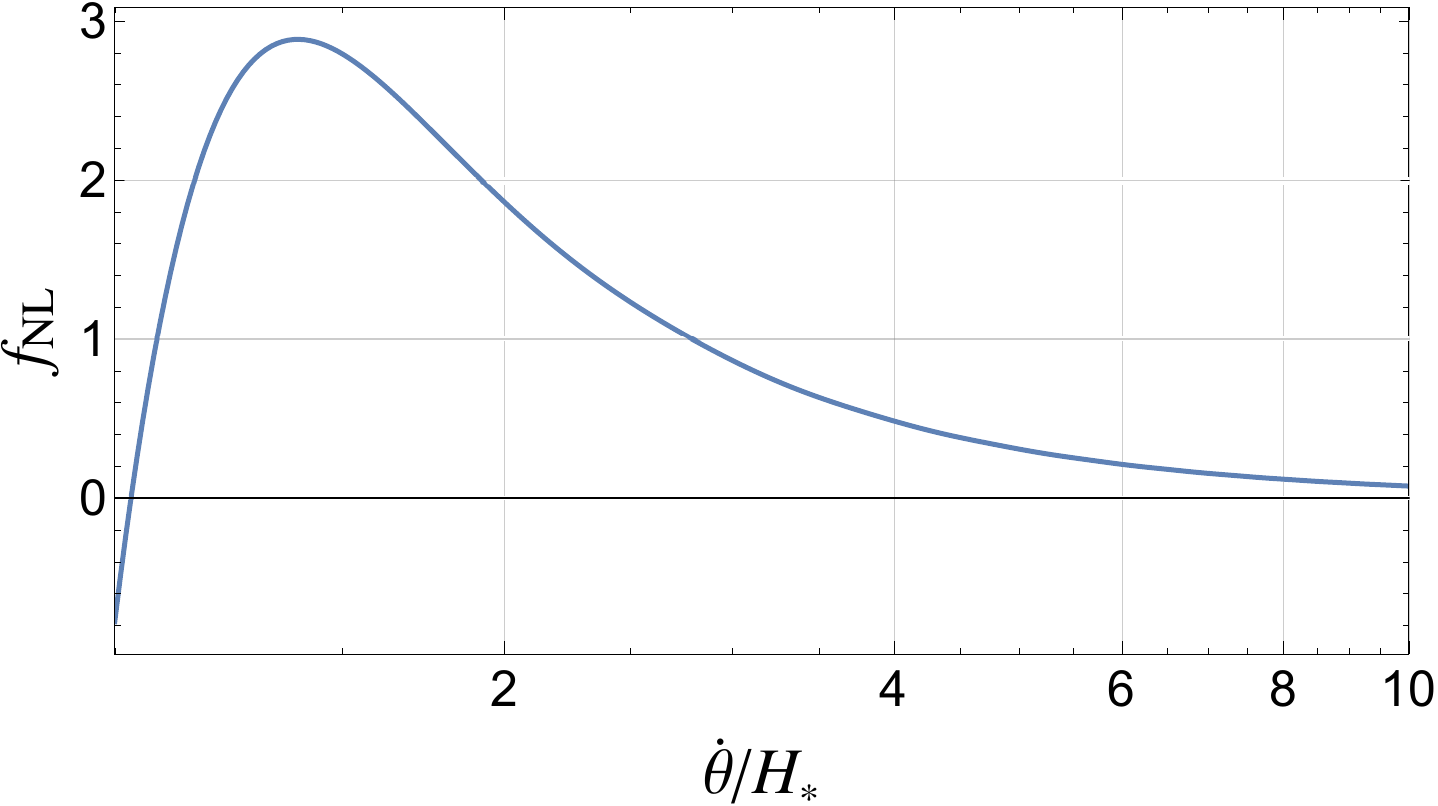}
	\end{center}
	\caption{\label{fig_eqfNL}
		The equilateral non-Gaussianity in the strong-mixing limit $\mu > H_\ast$ evaluated by the $H$-rescaling scheme for all exchange diagrams. 
	}
\end{figure}

A numerical estimation of the total non-Gaussianity in the equilateral limit ($k_1=k_2=k_3$) from all cubic interactions \eqref{cubic_interactions_Hami} is given in Fig. \ref{fig_eqfNL} with $\lambda = 0.01$ and $\Lambda = 2$. This result is evaluated by the $H$-rescaling scheme with the definition of $f_{NL}$ based on \cite{Assassi:2013gxa} as
\begin{align}\label{def:fNL}
    B^{eq}_\zeta = \left(\frac{H}{R_0\dot{\theta}_0}\right)^3 B^{eq}_\theta = \Delta_\zeta^4 \, f_{NL} \frac{18}{5}.
\end{align}
Note that the parameter $H$ in \eqref{def:fNL} is rescaled according to \eqref{H_rescaling} since we evaluate $P_\theta$ and $B_\theta\equiv (H/H_\ast)^3 B_\theta^\ast$ from the mode functions \eqref{mode EoM} with a reference parameter $H_\ast$. As a result, the bispectrum amplitude is
\begin{align}\nonumber
    f_{NL} &= \frac{5}{18} \Delta_\zeta^{-4}\left(\frac{\Delta_\zeta^2}{P_\theta^\ast} \right)^{3/2} f^{-3/2} \left(\frac{H_\ast}{H}\right)^3 B_\theta^{eq}, \\
    &= \frac{5}{18}\Delta_\zeta^{-1}f^{-3/2} I(t), 
    \label{fNL}
\end{align}
where $f$ and $I(t) = (2\pi)^3 B_\theta^\ast/H_\ast^3$ are computed by the EoM approach. The result \eqref{fNL} is independent of the choice of $\dot{\theta}_\ast$ or $H_\ast$.
\footnote{For the numerical estimation in Fig. \ref{fig_eqfNL}, we have applied the Wick rotation technique \cite{An:2017hlx,Chen:2017ryl} to mode functions to avoid the slow convergence in the UV limit.}  
The asymptotic value $f_{NL} \sim \mathcal{O}(10^{-2})$ in the limit $\dot{\theta}_0/H \gg 1$ is consistent with the estimation by integrating out the heavy Higgs field with an effective speed of sound \cite{Gong:2013sma}.

\smallskip
\noindent
\textbf{Non-analytic scaling.}
Away from the equilateral limit we can test the momentum scaling in bispectrum. In general, the particle exchange with a heavy field exhibits both local and non-local processes. These processes result in the analytic and non-analytic momentum scaling in bispectrum, respectively.   
%The oscillatory signatures in the bispectrum (such as in the right panel of Fig. \ref{fig_bispectrum}) is sourced by the non-analytic components of the mode functions that are not included in local operators.
The non-local process comes from components in the heavy-field mode functions which are oscillating in the late-time limit. 
 In other words, non-analytic signals do not appear if we integrate out the heavy field through the EFT approach from the beginning.\footnote{However, the leading non-analytic contribution can be captured if heavy-field operators are only partially integrated out \cite{Iyer:2017qzw}.}
For a weakly coupled system, the non-analytic components in the correlation functions can be computed explicitly by the in-in formalism \cite{Chen:2012ge,Lee:2016vti}. In the weak-mixing limit, the oscillatory feature in the bispectrum are generated by a heavy field with a mass $m/H > 3/2$, and they have the generic suppression factor $\sim e^{-\pi L}$ (with a canonical speed of sound $c_\pi^2 = 1$), where the scaling factor $L$ is given by \eqref{def:L}. 

\begin{figure}
	\begin{center}
		\includegraphics[width=100mm]{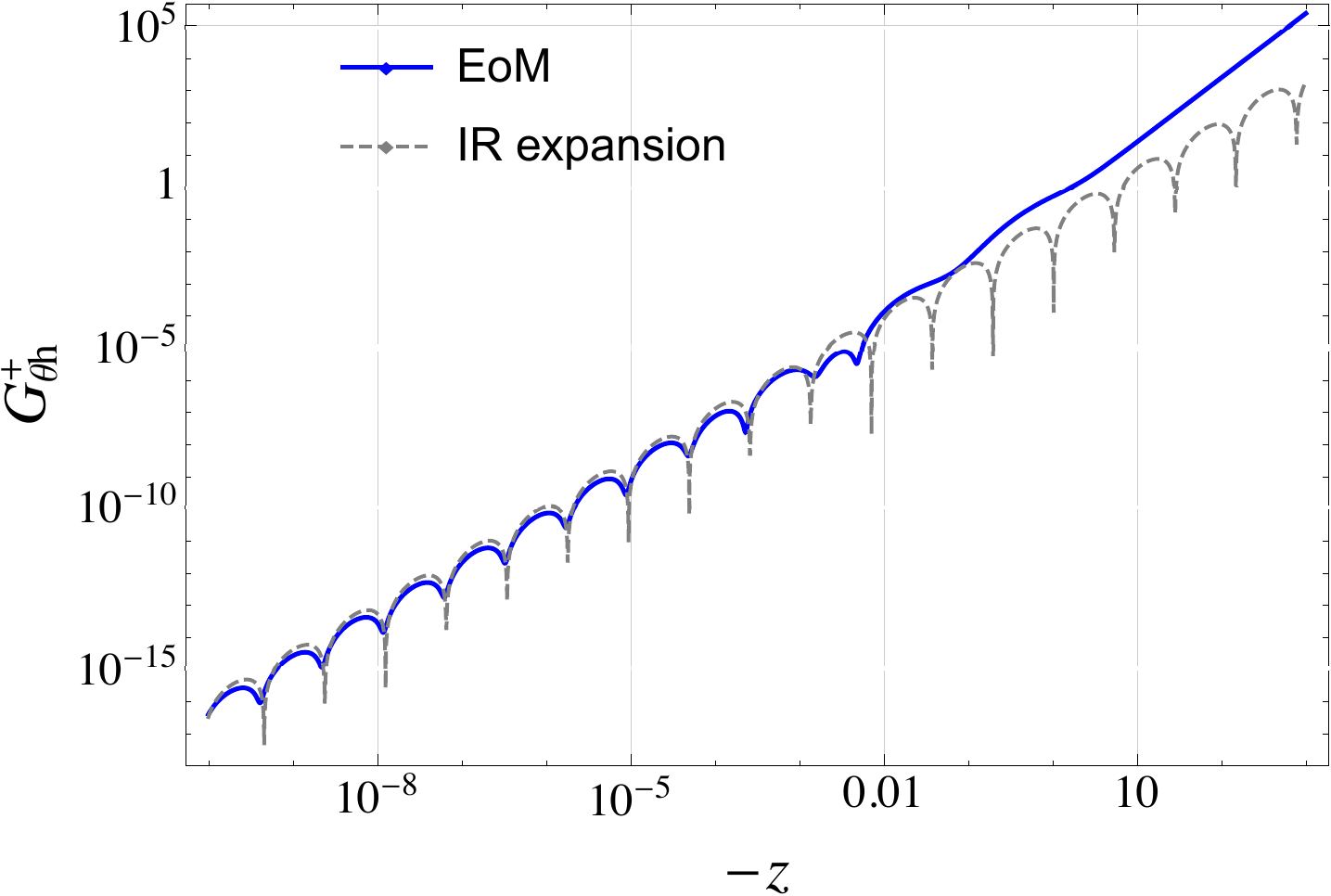}
	\end{center}
	\caption{\label{fig_non_analytic}
		The evolution of the component $G^+_{\theta h}$ in the mixed two-point function \eqref{two-pt mixed} with respect to $z = k\eta$ where the EoM approach is described in Appendix \ref{Appx_EoM} and the IR expansion is described in Appendix \ref{Appx_IR}. 
	}
\end{figure}

To see the non-analytic effect in the strongly coupled $\phi$-$h$ system, it is convenient to expand the mode function in the late-time limit $\vert\eta \vert\rightarrow 0$ where the approximated expression is found in \cite{An:2017hlx} as (see Appendix \ref{Appx_IR} for details)
\begin{align}\label{delta_h late_time}
\delta h_I \rightarrow \frac{H}{\sqrt{4k^3}}(-k\eta)^{3/2} \left[  B_+(-k\eta)^{i L_h} + B_-(-k\eta)^{-i L_h} +B_{\ast}(-k\eta)^{1/2} + \cdots \right].
\end{align}
Here $B_{\ast}$ and $B_\pm$ are coefficients that in general can only be fixed by fitting with numerical results, and the parameter $L_h$ is defined as
\begin{align}
L_h \equiv \sqrt{\frac{\mu_h^2}{H^2}-\frac{9}{4}} = \sqrt{\frac{m_h^2}{H^2c_h^2}-\frac{9}{4}}.
\end{align}
The imaginary powers $\pm i L_h$ in the late-time expansion \eqref{delta_h late_time} are the sources of the non-analytic scaling in correlation functions. In order to justify the late-time expansion \eqref{delta_h late_time}, we rewrite the mixed two-point function \eqref{eq:2-point mix} as
\begin{align}\label{two-pt mixed}
    \left\langle \delta\theta_{\bf k} (\eta) \delta h_{\bf q} (\eta) \right\rangle 
  =  (2\pi)^3 \delta^3({\bf k + q}) \left[G_{\theta h}^+(\eta) + G_{\theta h}^-(\eta)\right]
\end{align}
with $G_{\theta h}^\pm =  \delta\theta^{\pm}_k\delta h^{\pm\ast}_k = G^{\pm\ast}_{h\theta}$, where the two sets of independent mode functions are defined in \eqref{mode function: dtheta} and \eqref{mode function: dsigma}. 
By using the vanish of the equal time commutator
\begin{align}
    \left[\delta\theta_k(\eta),\delta h_k(\eta)\right] =
    G_{\theta h}^+ + G_{\theta h}^- - G_{h\theta }^+ - G_{h\theta }^- = 0,
\end{align}
we finds $\textrm{Im}[ G_{\theta h}^+ + G_{\theta h}^- ] =0$. Given that the late-time expansion of $\delta\theta$ is led by a constant, where following Appendix \ref{Appx_IR} we parametrize as $\delta\theta \rightarrow H A/ \sqrt{4k^3}$ with a phase $\vert A \vert = 1$, we may express in the limit $\eta \rightarrow 0$ that
\begin{align}\label{late-time Gtheta_h}
    G_{\theta h}^\pm \rightarrow \frac{H^2}{4k^2} (-k\eta)^{3/2} 
    \textrm{Re}\left[A B_+^\ast (-k\eta)^{-i L_h} + AB_-^\ast (-k\eta)^{i L_h}\right].
\end{align}
We use this expression for $G^+_{\theta h}$ to compare with that numerically solved by the EoM method in Fig. \ref{fig_non_analytic}. The results confirms that $L_h$ features the non-analytic scaling of the late-time mode functions, which is associated with the heavy Higgs scale $\mu_h$.

We remark that the oscillatory signature for the heavy particle production vanishes as $\mu_h/H < 3/2$ where $L_h$ becomes an imaginary number. In this case the non-analytic component has a non-integer scaling $(-k\eta)^{\Delta_\pm}$ with $\Delta = 3/2 \pm i L_h$. Enhancement of the non-Gaussian spectra due to $\Delta_+ < 3/2$ was investigated in \cite{An:2017rwo}.

%Recalling that the pure local-type non-Gaussianity is led by $L_h \rightarrow i 3/2$ (or $\mu_h \rightarrow 0$) while the equilateral-type is attributed to $L_h > 0$ (or $\mu_h > 3H/2$).

%\section{Observational impact of Heavy Higgs}

\begin{figure}
	\begin{center}
		\includegraphics[width=77mm]{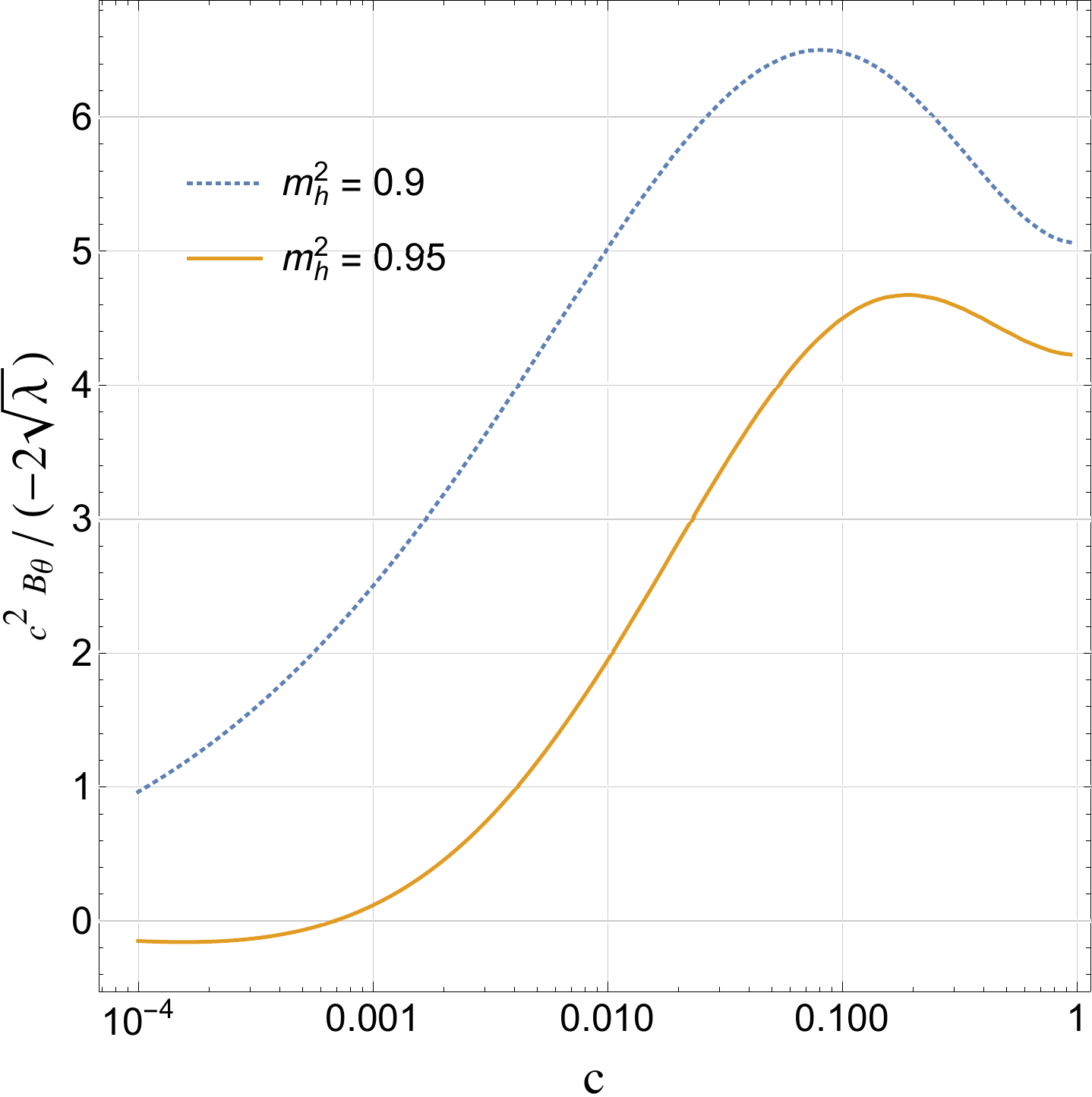}
		\hfill
		\includegraphics[width=78mm]{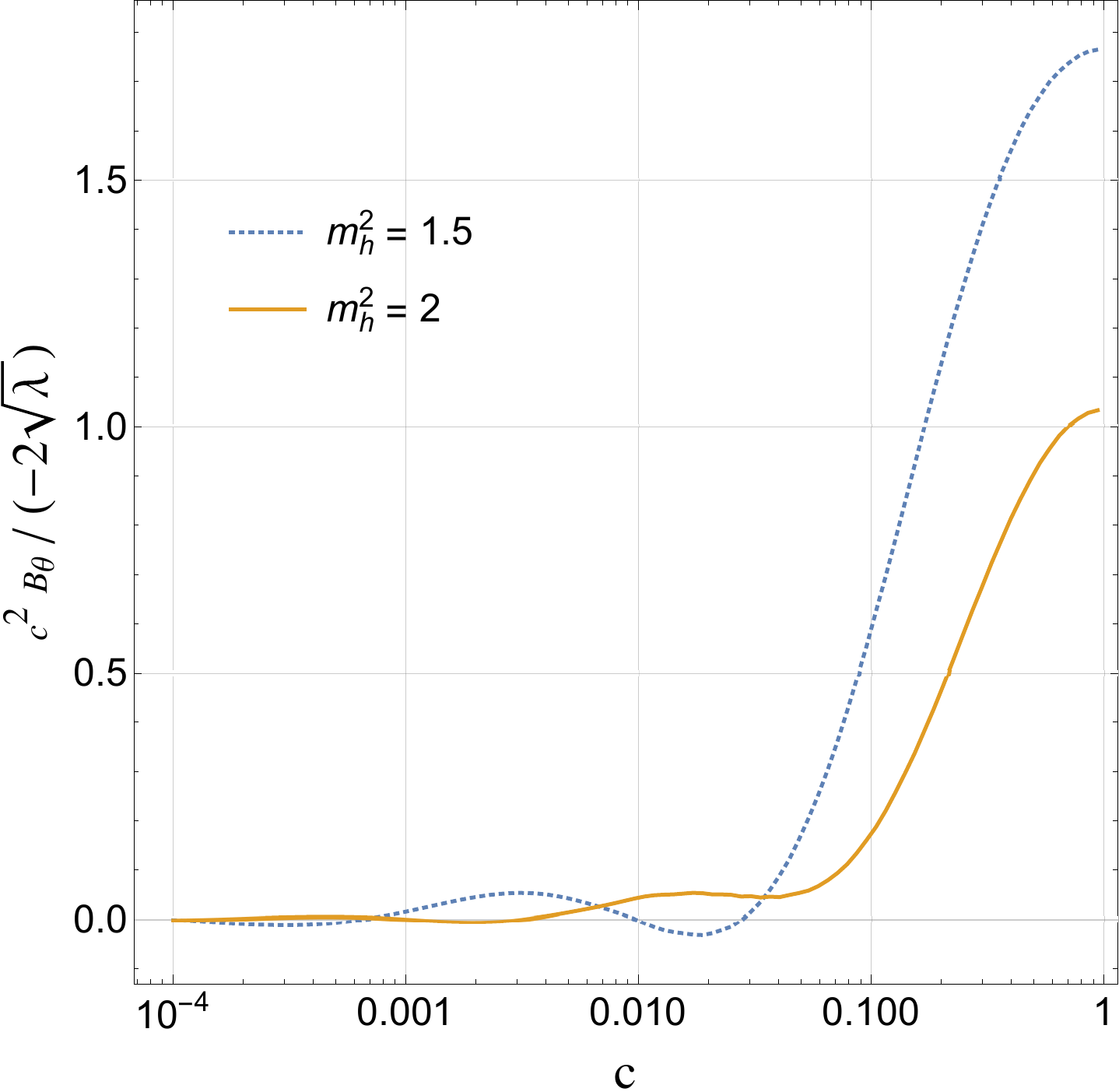}
	\end{center}
	\caption{\label{fig_bispectrum}
		The bispectrum $B_\theta$ due to Higgs-inflaton exchange processes of the intermediate-type (left panel) and the equilateral-type with oscillatory signatures (right panel). Parameters $\lambda =0.01$ and $\Lambda = 2$ are used, and $m_h$ is in Hubble unit.
	}
\end{figure}

\smallskip
\noindent
\textbf{Shape functions.}
A common classification of the shape function is based on the scaling behavior in the squeezed limit, for example, by taking $k_1 = k_2 = k$ and $k_3 = c k$ with $c \rightarrow 0$. For $c \ll 1$, a typical equilateral bispectrum scales as $S_\theta \sim c$ and a typical local bispectrum scales as $S_\theta \sim c^{-1}$  \cite{Chen:2010xka}, where the former peaks at the equilateral limit $c = 1$ and the later peaks at the squeezed limit $c =0 $. For a bispectrum scales as $S_\theta \sim c^{\nu}$ with $-1 < \nu < 1$ is referred to the intermediate shapes \cite{Chen:2009we,Chen:2009zp}. 
As an example, we plot in Fig. \ref{fig_bispectrum} the contribution from the interaction
\begin{align}
\tilde{\mathcal{H}}_{I,3} = a^3 \lambda h_0 \delta h_I^3,
\end{align}
with respect to different values of $m_h$ in Hubble unit and we have used the normalization $k/H = 1$. 
We can estimate the scaling of the triple exchange bispectrum in the squeezed limit by using the late-time expansion \eqref{late-time Gtheta_h} as
\begin{align}
    B_\theta \sim \frac{\lambda h_0}{H} c^{-3/2} \textrm{Im}
    \left[c^{-iL_h}\mathcal{I}_- + c^{iL_h}\mathcal{I}_+ \right],
\end{align}
where
\begin{align}
    \mathcal{I}_\pm = \int\frac{d\eta}{\eta^4} AB_\mp^\ast(-k\eta)^{3/2\pm i L_h}
    \left[G_{\theta h}^+(k\eta)+G_{\theta h}^-(k\eta)\right]^2.
\end{align}

The left panel of Fig. \ref{fig_bispectrum} shows a case with $\mu_h^2/H^2 < 9/4$ so that $L_h$ is imaginary and that $c^2 B_\theta \sim S_\theta \sim c^\nu$ with $0 < \nu < 1/2$. The bispectrum in these cases peak in between the equilateral and the squeezed limits (they can be referred as the quasi-equilateral shapes \cite{Chen:2009zp,Chen:2010xka}). The bispectra in the right panel of Fig. \ref{fig_bispectrum} have $\mu_h^2/H^2 > 9/4$, or $S_\theta \sim c^{1/2}\sin(L_h \ln c + \varphi)$ with a phase $\varphi$. They are basically of the equilateral shapes but they exhibit oscillatory signatures in the squeezed limit sourced by the imaginary part of $\nu$.
We remark that for a weak-mixing system the oscillatory signatures must be generated with $m_h^2/H^2 > 9/4$.
%imprints of the heavy Higgs production during inflation \cite{Arkani-Hamed:2015bza}.

\section{Summary and discussion}\label{Sec. Conclusion}

The heavy-lifting mechanism \cite{Kumar:2017ecc} is closely related to the cosmological collider research in the sense that one of its goals is to clarify the observability of SM signals in primordial non-Gaussianity. In general, the SM observability is improved with a broken gauge symmetry so that gauge fields (or at least the neutral bosons) can enter the late-time correlation functions via tree-level exchange with inflaton. It is nevertheless also interesting to consider observable SM signals from other possibilities. In any case, what can be more desirable is that the masses of SM fields can be lifted above the inflationary Hubble scale so that they can create non-analytic signals with non-integer or imaginary momentum scaling distinctive from that of the single-field inflation. 

In this work, we investigated a heavy-lifting scenario with broken gauge symmetry introduced by the slow-roll dynamics of inflation, characterized by the normalized inflaton velocity $\dot{\theta}$, where both the Higgs mass $m_h$ and the quadratic mixing $\mu$ are controlled by the background value $\dot{\theta}_0$. The coupled Higgs-inflaton system can be cast into a special type of quasi-single field inflation with a modified mass scale $\mu_h \equiv  (m_h^2 + \mu^2)^{1/2}$ which governs the non-analytic scaling behavior for the late-time mode functions. We have shown that Higgs indeed behaves as a heavy degree of freedom when $\mu_h > H$ and a strong-mixing $\mu > H$ can be realized without violating the loop expansion.   

After a proper rescaling of the power spectrum with respect to the observational constraint, especially when the system enters the strong-mixing regime, we confirmed that the non-Gaussianity in the equilateral limit in general can reach $f_{NL} \gtrsim \mathcal{O}(1)$. This value is compatible with current constraints from Cosmic Microwave Background observations \cite{Ade:2015ava}, and could be tested by the future Large Scale Structure or the 21-cm surveys \cite{Meerburg:2016zdz,MoradinezhadDizgah:2018ssw}. As the signature of heavy Higgs production, we demonstrated shape functions that exhibit non-analytic scaling in the squeezed limit. We have confirmed that the non-analytic scaling of the late-time correlation function is governed by the parameter $L_h$ with respect to the modified mass scale $\mu_h$.

It is remarkable that, even if the size of the Higgs signal is too weak to be detected in the large mass regime ($\mu_h \gg H$), the heavy Higgs field lifted by a strong-mixing can still result in an effective speed of sound $c_\theta^2 < 1$ for inflaton. Non-analytic signals sourced by other gauge fields are therefore relatively enhanced (or less suppressed) by the modified sound horizon with a small $c_\theta^2$ \cite{Lee:2016vti}. Although the specific system \eqref{system} considered in this work has a lower limit at $c_\theta^2 =1/3$, it is also straightforward to seek for the realization of $c_\theta^2 \ll 1$ from extended scenario in the full expression of the effective theory \eqref{L_phi_h}.

A rearrangement of the SM mass spectrum with the modified Higgs scale invoked by heavy-lifting is an important subject to be clarified, which has been postponed as a future study. In a very optimistic scenario where signals of more than one heavy-lifted gauge fields can be detected, it might be possible to test the renormalization group running of SM couplings up to the energy scale of inflation \cite{Chen:2016hrz}. These results might reveal some hints connected to the Naturalness of SM \cite{Kumar:2017ecc}, or the issue of Higgs instability \cite{EliasMiro:2011aa,Degrassi:2012ry,Buttazzo:2013uya}.

\section*{acknowledgments}
We thank Alexander Kusenko, Hayden Lee, Shao-Jiang Wang, Yi Wang, Zhong-Zhi Xianyu, Masahide Yamaguchi, Jun'ichi Yokoyama and Siyi Zhou for many helpful comments and inspiring discussions.
The author is grateful to Xingang Chen for his careful reading of the manuscript and Louis Yang for his contribution at the early stage of this work.
YPW is supported by JSPS International Research Fellows and JSPS KAKENHI Grant-in-Aid for Scientific Research No. 17F17322.

\appendix

\section{The equation of motion method}\label{Appx_EoM}
In this appendix we describe the formulation of the EoM method used in Section \ref{Sec. imprints}.
According to the first-principles of the in-in formalism \cite{Chen:2010xka,Wang:2013eqj,Weinberg:2005vy}, the evolution of field perturbations are governed by the part of Hamiltonian $\tilde{H} = \tilde{H}[\delta\theta, \delta h, p_\theta, p_h; t]$ consisted by quadratic and higher-orders in perturbations, where $p_\theta$ and $p_h$ are conjugate momenta for $\theta$ and $h$. In general, one can perform the decomposition
\begin{align}
\tilde{H}[\delta\theta, \delta h, p_\theta, p_h; t] = H_0[\delta\theta, \delta h, p_\theta, p_h; t] + H_I[\delta\theta, \delta h, p_\theta, p_h; t], \nonumber
\end{align}
where $H_0$ includes only quadratic terms without $\phi$-$h$ interactions. $H_0$ is usually used to define the evolution of free fields $\delta\theta_I$ and $\delta h_I$  in the interaction picture. Since $\phi$-$h$ interactions are cast into $H_I$ as perturbations, mode functions of $\delta\theta_I$ and $\delta h_I$ must have vanished correlations, namely $\langle \delta\theta^I_{\bf k}\delta h^I_{\bf q}\rangle = 0$.

In a strongly coupled system, the quadratic interactions in $H_I$ can also play an important role from the initial time. Therefore, instead of defining mode functions with respect to $\delta\theta_I$ and $\delta h_I$, we impose the definition \cite{Chen:2015dga}:
\begin{align} 
\label{mode function: dtheta}
\delta\theta_{\bf k} &= 
\delta\theta^{+}_{ k} \hat{a}_{\bf k} + \delta\theta^{+\ast}_{ k} \hat{a}^\dagger_{\bf -k} + \delta\theta^{-}_{ k}\hat{b}_{\bf k} + \delta\theta^{-\ast}_{ k} \hat{b}^\dagger_{\bf -k}, \\\nonumber
\label{mode function: dsigma}
\delta h_{\bf k} & =
\delta h^{+}_{ k} \hat{a}_{\bf k} + \delta h^{+\ast}_{ k} \hat{a}^\dagger_{\bf -k} + \delta h^{-}_{ k}\hat{b}_{\bf k} + \delta h^{-\ast}_{ k} \hat{b}^\dagger_{\bf -k}, \\
p_{\theta, \bf k} &= 
p_{\theta,  k}^{+} \hat{a}_{\bf k} + p_{\theta,  k} ^{+\ast} \hat{a}^\dagger_{\bf -k} + p_{\theta,  k}^{-}\hat{b}_{\bf k} + p_{\theta,  k} ^{-\ast} \hat{b}^\dagger_{\bf -k}, 
\\\nonumber
%\label{mode function: dsigma}
p_{h, \bf k} & =
p_{h,  k}^{+} \hat{a}_{\bf k} + p_{h,  k} ^{+\ast} \hat{a}^\dagger_{\bf -k} + p_{h,  k}^{-}\hat{b}_{\bf k} + p_{h,  k} ^{-\ast} \hat{b}^\dagger_{\bf -k},
\end{align}
which satisfies the canonical commutation relation at initial time $t_0$ as
\begin{align}
[\delta\theta(\textbf{x},t_0), p_\theta(\textbf{y},t_0)] = [\delta h(\textbf{x},t_0), p_h(\textbf{y},t_0)] &= i \delta^3(\textbf{x}-\textbf{y}), \\\nonumber
[\delta\theta(\textbf{x},t_0), \delta\theta(\textbf{y},t_0)] = [\delta h(\textbf{x},t_0), \delta h(\textbf{y},t_0)]  
&= [\delta\theta(\textbf{x},t_0), \delta h(\textbf{y},t_0)] = 0, \\\nonumber
[p_\theta(\textbf{x},t_0), p_\theta(\textbf{y},t_0)] = [p_h(\textbf{x},t_0), p_h(\textbf{y},t_0)] &=[p_\theta(\textbf{x},t_0), p_h(\textbf{y},t_0)] =0.
\end{align}
This also implies the mutual operators shall follow
\begin{align}
[\hat{a}_{\bf k}, \hat{a}^\dagger_{\bf -q}] &= [\hat{b}_{\bf k}, \hat{b}^\dagger_{\bf -q}] 
= (2\pi)^3 \delta^3( {\bf k + q}), \\\nonumber
[\hat{a}_{\bf k}, \hat{a}_{\bf -q}] = [\hat{b}_{\bf k}, \hat{b}_{\bf -q}] 
&= [\hat{a}_{\bf k}, \hat{b}_{\bf -q}] = [\hat{a}^\dagger_{\bf k}, \hat{b}^\dagger_{\bf -q}]  
= [\hat{a}_{\bf k}, \hat{b}^\dagger_{\bf -q}] = [\hat{b}_{\bf k}, \hat{a}^\dagger_{\bf -q}] = 0.
\end{align}
The two-point correlation function is led by contributions from all vacuum as
\begin{align}\label{eq:2-point theta}
\left\langle \delta\theta_{\bf k} (z) \delta\theta_{\bf q} (z) \right\rangle =
(2\pi)^3 \delta^3({\bf k + q}) \left[\delta\theta^+_{  k}(z) \delta\theta^{+ \ast}_{  q} (z) + \delta\theta^-_{  k}(z) \delta\theta^{-\ast}_{  q}(z) \right], \\
\left\langle \delta h_{\bf k} (z) \delta h_{\bf q} (z) \right\rangle =
(2\pi)^3 \delta^3({\bf k + q}) \left[\delta h^+_{  k}(z) \delta h^{+ \ast}_{  q} (z) + \delta h^-_{  k}(z) \delta h^{-\ast}_{  q}(z) \right].
\end{align}  
It is important to note that the definition \eqref{mode function: dtheta} and \eqref{mode function: dsigma} admits the non-vanished correlation
\begin{align}\label{eq:2-point mix}
  \left\langle \delta\theta_{\bf k} (z) \delta h_{\bf q} (z) \right\rangle 
  =  (2\pi)^3 \delta^3({\bf k + q}) \left[\delta \theta^+_{  k}(z) \delta h^{+ \ast}_{  q} (z) + \delta \theta^-_{  k}(z) \delta h^{-\ast}_{  q}(z)\right].
\end{align}

The non-diagonal mode functions defined in \eqref{mode function: dtheta} can be taken as modified ``free fields'' for computing higher-order perturbations in the interaction picture (although they are indeed strongly-coupled). In other words, we are in fact performing the decomposition of $\tilde{H}$ as
\begin{align}\label{EoM_dec_modified}
\tilde{H}[\delta\theta, \delta h, p_\theta, p_h; t] = \tilde{H}_0[\delta\theta, \delta h, p_\theta, p_h; t] + \tilde{H}_I[\delta\theta, \delta h, p_\theta, p_h; t], 
\end{align}
where $\tilde{H}_0$ includes all quadratic terms that give the coupled EoM \eqref{EoM:coupled} and $\tilde{H}_I$ contains only cubic or higher-order in perturbations.

Let us derive $\tilde{H}_0$ and $\tilde{H}_I$ for the $\phi$-$h$ system given by \eqref{quadratic perturbations canonical} and \eqref{cubic interactions}. The conjugate momenta are
\begin{align}\nonumber
p_\theta &= \frac{\partial \mathcal{L}}{\partial \delta\dot{\theta}_c} = 
     \delta\dot{\theta}_c + \mu \delta h + 2 \frac{h_0}{R_0^2}\delta\dot{\theta}_c \delta h + \frac{\dot{\theta}_c}{R_0^2} \delta h^2, \\
p_h &=  \frac{\partial \mathcal{L}}{\partial \delta \dot{h}} = \delta \dot{h},
\end{align} 
where $\mathcal{L} = \mathcal{L}_2 +\mathcal{L}_3 $. Replacing $\delta\dot{\theta}_c$ and $\delta \dot{h}$ by $p_\theta$ and $p_h$ while keeping linear in perturbations, we can obtain the modified interaction-picture fields via $\delta\dot{\theta}_I = \partial \tilde{\mathcal{H}}_0/\partial p_\theta$ and $\delta\dot{h}_I = \partial \tilde{\mathcal{H}}_0/\partial p_h$. This results in
\begin{align}
\tilde{\mathcal{H}}_0 = \frac{1}{2} \left[\delta\dot{\theta}_I^2 + \frac{1}{a^2}(\partial_i \delta\theta_I)^2 +\delta\dot{h}_I^2 + \frac{1}{a^2}(\partial_i \delta h_I)^2 + (m_h^2 + \mu^2) \delta h_I^2 - 2\mu \delta\dot{\theta}_I \delta h_I \right],
\end{align} 
as well as the cubic interactions $\tilde{\mathcal{H}}_{I,3}$ of the form
\begin{align}\nonumber
\tilde{\mathcal{H}}_{I,3}^{(1)} &= -\frac{h_0}{R_0^2} \delta\dot{\theta}_I^2 \delta h_I + \frac{h_0}{a^2R_0^2} (\partial_i \delta\theta_I)^2 \delta h_I,\\ \label{cubic_interactions_Hami}
\tilde{\mathcal{H}}_{I,3}^{(2)} &= \left(2\frac{h_0}{R_0^2}\mu - \frac{\dot{\theta}_c}{R_0^2}\right) \delta\dot{\theta}_I \delta h_I^2,\\\nonumber
\tilde{\mathcal{H}}_{I,3}^{(3)} &= \left(\lambda h_0 -\frac{h_0}{R_0^2}\mu^2 + \frac{\dot{\theta}_c}{R_0^2}\mu \right)\delta h_I^3.
\end{align}
They are classified by the single, double and triple exchange vertices in Fig. \ref{fig_Feynmen diagrams}, respectively. 

To determine the mode functions at the initial time, let us rewrite the equations of motion \eqref{eom: theta-h} with respect to the dimensionless time variable $z = k \eta$, which read
\begin{align}\label{EoM:coupled}
\frac{\partial^2}{\partial z^2}\delta\theta -\frac{2}{z}\frac{\partial }{\partial z}\delta\theta +\delta\theta 
&= \frac{\mu}{H} \left(\frac{\partial }{z \partial z}\delta h - 3\frac{\delta h}{z^2}\right), \\ \nonumber
\frac{\partial^2}{\partial z^2}\delta h-\frac{2}{z}\frac{\partial }{\partial z}\delta h +\left(1+ \frac{m_h^2}{H^2z^2}\right)\delta h
&= - \frac{\mu}{H} \frac{\partial }{z \partial z}\delta \theta.
\end{align}
In the limit of $-z \gg 1$, one can drop terms in proportion to $z^{-2}$ and solve the perturbations by the factorization \cite{An:2017hlx} of the form
\begin{align}
\delta\theta = \frac{H}{\sqrt{4k^3}}A e^{-i z}, \quad \delta h = \frac{H}{\sqrt{4k^3}}B e^{-i z},
\end{align}
to obtain $A = (-z)^{1\pm i \mu/(2H)}$ and $B = \pm i A$. These results give the initial states \eqref{mode EoM} for the numerical computations.

\section{The late-time expansion}\label{Appx_IR}
In this appendix we perform power series expansion of the mode functions to match their coefficients in the late-time limit $z\rightarrow 0$.
A general power series expansion of the solutions for the equations of motion \eqref{EoM:coupled} is given by \cite{An:2017rwo} as
\begin{align}
    \delta\theta_k^\pm &= \frac{H}{\sqrt{4k^3}}\sum_{n,\nu} A_{n\nu}^\pm (-z)^{n+\nu},\\
    \delta h_k^\pm &= \frac{H}{\sqrt{4k^3}}\sum_{n,\nu} B_{n\nu}^\pm (-z)^{n+\nu},
\end{align}
where $n$ is summed from $0$ to $\infty$. The branch value $\nu$ can be found from the $n = 0$ equations in terms of the power series for the vanish of $(-z)^{\nu - 2}$ as
\begin{align}\label{constraint:n=0}
    \left(\nu A_{0\nu}- \frac{\mu}{H}B_{0\nu} \right) (\nu - 3) &= 0, \\\nonumber
    \left[B_{0\nu}(\nu - 3) + \frac{\mu}{H}A_{0\nu}\right]\nu + \frac{m_h^2}{H^2}B_{0\nu} &= 0.
\end{align}
There are four possible branch values given by
\begin{align}
    \nu = 0, \; 3, \; \frac{3}{2}\pm i L_h.
\end{align}
The $n = 1$ equations for the vanish of $(-z)^{\nu - 1}$ are
\begin{align}
    \left[(\nu+1) A_{1\nu}- \frac{\mu}{H}B_{1\nu} \right] (\nu - 2) &= 0, \\\nonumber
    \left[B_{1\nu}(\nu - 2) + \frac{\mu}{H}A_{1\nu}\right](\nu + 1) + \frac{m_h^2}{H^2}B_{1\nu} &= 0.
\end{align}
The $n \geq 2$ equations for the vanish of $(-z)^{\nu - 2 + n}$ are
\begin{align}
    \left[(\nu+n) A_{n\nu}- \frac{\mu}{H}B_{n\nu} \right] (\nu - 3+n) +A_{n\nu} &= 0, \\\nonumber
    \left[B_{n\nu}(\nu - 3 + n) + \frac{\mu}{H}A_{n\nu}\right](\nu + n) + \frac{m_h^2}{H^2}B_{n\nu} + B_{n\nu}&= 0.
\end{align}
Note that odd-$n$ coefficients can be set to zero to remove redundant solutions. 
In the late-time limit $z\rightarrow 0$, the mode functions (for both $\pm$ states of \eqref{mode function: dtheta} and \eqref{mode function: dsigma}) are led by
\begin{align}
    \delta\theta_k &= \frac{H}{\sqrt{4k^3}}(-z)^{3/2}\left[A_{00}(-z)^{-3/2}+A_{0+}(-z)^{iL_h} + A_{0-}(-z)^{-iL_h}+\cdots\right],\\
    \delta h_k &= \frac{H}{\sqrt{4k^3}}(-z)^{3/2}\left[B_{0+}(-z)^{iL_h} + B_{0-}(-z)^{-iL_h}+ B_{20}(-z)^{1/2}\cdots\right],
\end{align}
where $B_{00} = 0$ due to the constraint of \eqref{constraint:n=0}. One can find the relation between coefficients from \eqref{constraint:n=0}, where $(\frac{3}{2}\pm i L_h)A_{0\pm} =\mu B_{0\pm}/H $.

The two non-integer branch values $v = \frac{3}{2}\pm i L_h$ are in fact solved from the part of equation $\nu A_{0\nu} = \mu B_{0\nu}/H$ in \eqref{constraint:n=0}. These solutions imply a possible late-time approximation of the theory \eqref{quadratic perturbations canonical} as
\begin{align}\label{late-time quadratic perturbations}
    \mathcal{L}_2 \approx \frac{1}{2}\left[ \delta\dot{\theta}^2_c + \delta\dot{h}^2 - \frac{1}{2a^2}(\partial_i \delta h)^2 - m_h^2\delta h^2 + 2\mu \delta h \delta\dot{\theta}_c\right],
\end{align}
which results in the effective relation $\delta h = -\delta\dot{\theta}_c/\mu$. 
Putting this effective relation back into \eqref{late-time quadratic perturbations}, one obtain the Lagrangian only for $\delta h$, which gives the familiar equation of motion as
\begin{align}
    \frac{\partial^2}{\partial z^2}\delta h -\frac{2}{z}\frac{\partial }{\partial z}\delta h +\left( \frac{m_h^2 + \mu^2}{H^2 z^2}\right)\delta h =0.
\end{align}
Assuming the usual Bunch-Davies vacuum, the mode function of the equation of motion is well-known:
\begin{align}\label{late-time mode function}
    \delta h = H \sqrt{\frac{\pi}{4k^3}}(-z)^{3/2}e^{i(iL_h+1/2)\pi/2} H^{(1)}_{i L_h}(-z),
\end{align}
where $H^{(1)}_\nu$ is Hankel function of the first kind.
Expanding the Hankel function at $z \rightarrow 0$, we find that
\begin{align}
    \delta h \rightarrow -\frac{i}{\pi} e^{i\pi/4} & H \sqrt{\frac{\pi}{4k^3}} (-z)^{3/2} \\\nonumber
    \times & \left[e^{\pi L_h/2}\Gamma(-i L_h)\left(-\frac{z}{2}\right)^{i L_h}+ e^{-\pi L_h/2}\Gamma(i L_h)\left(-\frac{z}{2}\right)^{-i L_h}\right].
\end{align}
Therefore it is easy to match the leading coefficients in the late-time expansion up to a phase difference as
\begin{align}\label{late-time B}
    \vert B_{0\pm} \vert = \left| e^{\pm\pi L_h/2} \Gamma(\mp i L_h)/\sqrt{\pi}\right|.
\end{align}
A comparison of the late-time expansion given by \eqref{late-time B} with the numerical result of the EoM approach is shown in Fig. \ref{fig_non_analytic}. 
We remark that although the power series expansion was first derived in Ref. \cite{An:2017rwo} for studying the cases with $m_h^2+\mu^2 < 9H^2/4$, the late-time effective theory \eqref{late-time quadratic perturbations} still holds in cases with $m_h^2+\mu^2 > 9H^2/4$ since the approximation does not rely on the value of the parameter $m_h$ nor $\mu$.


\begin{thebibliography}{99}

\bibitem{Baumann:2014nda} 
D.~Baumann and L.~McAllister,
``Inflation and String Theory,''
arXiv:1404.2601 [hep-th].

%\cite{Copeland:1994vg}
\bibitem{Copeland:1994vg} 
  E.~J.~Copeland, A.~R.~Liddle, D.~H.~Lyth, E.~D.~Stewart and D.~Wands,
  %``False vacuum inflation with Einstein gravity,''
  Phys.\ Rev.\ D {\bf 49}, 6410 (1994)
  doi:10.1103/PhysRevD.49.6410
  [astro-ph/9401011].
  %%CITATION = doi:10.1103/PhysRevD.49.6410;%%
  %1000 citations counted in INSPIRE as of 20 Nov 2018



%\cite{Yamaguchi:2011kg}
\bibitem{Yamaguchi:2011kg} 
M.~Yamaguchi,
``Supergravity based inflation models: a review,''
Class.\ Quant.\ Grav.\  {\bf 28}, 103001 (2011)
%doi:10.1088/0264-9381/28/10/103001
[arXiv:1101.2488 [astro-ph.CO]].
%%CITATION = doi:10.1088/0264-9381/28/10/103001;%%
%91 citations counted in INSPIRE as of 01 Nov 2018

%\cite{Baumann:2011nk}
\bibitem{Baumann:2011nk} 
D.~Baumann and D.~Green,
``Signatures of Supersymmetry from the Early Universe,''
Phys.\ Rev.\ D {\bf 85}, 103520 (2012)
%doi:10.1103/PhysRevD.85.103520
[arXiv:1109.0292 [hep-th]].
%%CITATION = doi:10.1103/PhysRevD.85.103520;%%
%146 citations counted in INSPIRE as of 01 Nov 2018

%\cite{Assassi:2013gxa}
\bibitem{Assassi:2013gxa} 
  V.~Assassi, D.~Baumann, D.~Green and L.~McAllister,
  %``Planck-Suppressed Operators,''
  JCAP {\bf 1401}, 033 (2014)
  doi:10.1088/1475-7516/2014/01/033
  [arXiv:1304.5226 [hep-th]].
  %%CITATION = doi:10.1088/1475-7516/2014/01/033;%%
  %52 citations counted in INSPIRE as of 06 Dec 2018

\bibitem{Chen:2009we} 
X.~Chen and Y.~Wang,
``Large non-Gaussianities with Intermediate Shapes from Quasi-Single Field Inflation,''
Phys.\ Rev.\ D {\bf 81}, 063511 (2010)
%  doi:10.1103/PhysRevD.81.063511
[arXiv:0909.0496 [astro-ph.CO]].

\bibitem{Chen:2009zp} 
X.~Chen and Y.~Wang,
``Quasi-Single Field Inflation and Non-Gaussianities,''
JCAP {\bf 1004}, 027 (2010)
%doi:10.1088/1475-7516/2010/04/027
[arXiv:0911.3380 [hep-th]].

%\cite{Noumi:2012vr}
\bibitem{Noumi:2012vr} 
T.~Noumi, M.~Yamaguchi and D.~Yokoyama,
``Effective field theory approach to quasi-single field inflation and effects of heavy fields,''
JHEP {\bf 1306}, 051 (2013)
%doi:10.1007/JHEP06(2013)051
[arXiv:1211.1624 [hep-th]].
%%CITATION = doi:10.1007/JHEP06(2013)051;%%
%83 citations counted in INSPIRE as of 05 May 2018

\bibitem{Gong:2013sma}
J.~O.~Gong, S.~Pi and M.~Sasaki,
``Equilateral non-Gaussianity from heavy fields,''
JCAP {\bf 1311} (2013) 043
[arXiv:1306.3691 [hep-th]].

\bibitem{Arkani-Hamed:2015bza} 
N.~Arkani-Hamed and J.~Maldacena,
``Cosmological Collider Physics,''
arXiv:1503.08043 [hep-th].

%\cite{Kehagias:2015jha}
\bibitem{Kehagias:2015jha} 
  A.~Kehagias and A.~Riotto,
  %``High Energy Physics Signatures from Inflation and Conformal Symmetry of de Sitter,''
  Fortsch.\ Phys.\  {\bf 63}, 531 (2015)
  doi:10.1002/prop.201500025
  [arXiv:1501.03515 [hep-th]].
  %%CITATION = doi:10.1002/prop.201500025;%%
  %25 citations counted in INSPIRE as of 21 Nov 2018

%\cite{Kehagias:2017cym}
\bibitem{Kehagias:2017cym} 
  A.~Kehagias and A.~Riotto,
  %``On the Inflationary Perturbations of Massive Higher-Spin Fields,''
  JCAP {\bf 1707}, no. 07, 046 (2017)
  doi:10.1088/1475-7516/2017/07/046
  [arXiv:1705.05834 [hep-th]].
  %%CITATION = doi:10.1088/1475-7516/2017/07/046;%%
  %23 citations counted in INSPIRE as of 21 Nov 2018

%\cite{Franciolini:2017ktv}
\bibitem{Franciolini:2017ktv} 
  G.~Franciolini, A.~Kehagias and A.~Riotto,
  %``Imprints of Spinning Particles on Primordial Cosmological Perturbations,''
  JCAP {\bf 1802}, no. 02, 023 (2018)
  doi:10.1088/1475-7516/2018/02/023
  [arXiv:1712.06626 [hep-th]].
  %%CITATION = doi:10.1088/1475-7516/2018/02/023;%%
  %19 citations counted in INSPIRE as of 21 Nov 2018

%\cite{Meerburg:2016zdz}
\bibitem{Meerburg:2016zdz} 
P.~D.~Meerburg, M.~Münchmeyer, J.~B.~Muñoz and X.~Chen,
``Prospects for Cosmological Collider Physics,''
JCAP {\bf 1703}, no. 03, 050 (2017)
%doi:10.1088/1475-7516/2017/03/050
[arXiv:1610.06559 [astro-ph.CO]].
%%CITATION = doi:10.1088/1475-7516/2017/03/050;%%
%17 citations counted in INSPIRE as of 14 Apr 2018

%\cite{Lee:2016vti}
\bibitem{Lee:2016vti} 
H.~Lee, D.~Baumann and G.~L.~Pimentel,
``Non-Gaussianity as a Particle Detector,''
JHEP {\bf 1612}, 040 (2016)
%doi:10.1007/JHEP12(2016)040
[arXiv:1607.03735 [hep-th]].
%%CITATION = doi:10.1007/JHEP12(2016)040;%%
%52 citations counted in INSPIRE as of 09 Oct 2018

\bibitem{Dimastrogiovanni:2015pla}
E.~Dimastrogiovanni, M.~Fasiello and M.~Kamionkowski,
``Imprints of Massive Primordial Fields on Large-Scale Structure,''
JCAP {\bf 1602}, 017 (2016).
[arXiv:1504.05993 [astro-ph.CO]].

\bibitem{Schmidt:2015xka}
F.~Schmidt, N.~E.~Chisari and C.~Dvorkin,
``Imprint of inflation on galaxy shape correlations,''
JCAP {\bf 1510}, no. 10, 032 (2015).
[arXiv:1506.02671 [astro-ph.CO]].


\bibitem{MoradinezhadDizgah:2018ssw} 
  A.~Moradinezhad Dizgah, H.~Lee, J.~B.~Munoz and C.~Dvorkin,
  ``Galaxy Bispectrum from Massive Spinning Particles,''
  arXiv:1801.07265 [astro-ph.CO].

%\cite{Saito:2018omt}
\bibitem{Saito:2018omt} 
R.~Saito and T.~Kubota,
``Heavy Particle Signatures in Cosmological Correlation Functions with Tensor Modes,''
JCAP {\bf 1806}, no. 06, 009 (2018)
%doi:10.1088/1475-7516/2018/06/009
[arXiv:1804.06974 [hep-th]].
%%CITATION = doi:10.1088/1475-7516/2018/06/009;%%
%3 citations counted in INSPIRE as of 09 Oct 2018

%\cite{Wang:2018tbf}
\bibitem{Wang:2018tbf} 
  Y.~Wang, Y.~P.~Wu, J.~Yokoyama and S.~Zhou,
  ``Hybrid Quasi-Single Field Inflation,''
  JCAP {\bf 1807}, no. 07, 068 (2018)
  %doi:10.1088/1475-7516/2018/07/068
  [arXiv:1804.07541 [astro-ph.CO]].
  %%CITATION = doi:10.1088/1475-7516/2018/07/068;%%

%\cite{Kumar:2018jxz}
\bibitem{Kumar:2018jxz} 
S.~Kumar and R.~Sundrum,
%``Seeing Higher-Dimensional Grand Unification In Primordial Non-Gaussianities,''
arXiv:1811.11200 [hep-ph].
%%CITATION = ARXIV:1811.11200;%%

%\cite{Goon:2018fyu}
\bibitem{Goon:2018fyu} 
G.~Goon, K.~Hinterbichler, A.~Joyce and M.~Trodden,
%``Shapes of gravity: Tensor non-Gaussianity and massive spin-2 fields,''
arXiv:1812.07571 [hep-th].
%%CITATION = ARXIV:1812.07571;%%

%\cite{Arkani-Hamed:2018kmz}
\bibitem{Arkani-Hamed:2018kmz} 
N.~Arkani-Hamed, D.~Baumann, H.~Lee and G.~L.~Pimentel,
``The Cosmological Bootstrap: Inflationary Correlators from Symmetries and Singularities,''
arXiv:1811.00024 [hep-th].
%%CITATION = ARXIV:1811.00024;%%

%\cite{Chen:2016uwp}
\bibitem{Chen:2016uwp} 
X.~Chen, Y.~Wang and Z.~Z.~Xianyu,
``Standard Model Background of the Cosmological Collider,''
Phys.\ Rev.\ Lett.\  {\bf 118}, no. 26, 261302 (2017)
%doi:10.1103/PhysRevLett.118.261302
[arXiv:1610.06597 [hep-th]].
%%CITATION = doi:10.1103/PhysRevLett.118.261302;%%
%17 citations counted in INSPIRE as of 01 Nov 2018

%\cite{Chen:2016hrz}
\bibitem{Chen:2016hrz} 
X.~Chen, Y.~Wang and Z.~Z.~Xianyu,
``Standard Model Mass Spectrum in Inflationary Universe,''
JHEP {\bf 1704}, 058 (2017)
%doi:10.1007/JHEP04(2017)058
[arXiv:1612.08122 [hep-th]].
%%CITATION = doi:10.1007/JHEP04(2017)058;%%
%13 citations counted in INSPIRE as of 01 Nov 2018


%\cite{EliasMiro:2011aa}
\bibitem{EliasMiro:2011aa} 
J.~Elias-Miro, J.~R.~Espinosa, G.~F.~Giudice, G.~Isidori, A.~Riotto and A.~Strumia,
``Higgs mass implications on the stability of the electroweak vacuum,''
Phys.\ Lett.\ B {\bf 709}, 222 (2012)
%doi:10.1016/j.physletb.2012.02.013 
[arXiv:1112.3022 [hep-ph]].
%%CITATION = doi:10.1016/j.physletb.2012.02.013;%%
%371 citations counted in INSPIRE as of 03 May 2018

%\cite{Degrassi:2012ry}
\bibitem{Degrassi:2012ry} 
G.~Degrassi, S.~Di Vita, J.~Elias-Miro, J.~R.~Espinosa, G.~F.~Giudice, G.~Isidori and A.~Strumia,
``Higgs mass and vacuum stability in the Standard Model at NNLO,''
JHEP {\bf 1208}, 098 (2012)
%doi:10.1007/JHEP08(2012)098
[arXiv:1205.6497 [hep-ph]].
%%CITATION = doi:10.1007/JHEP08(2012)098;%%
%982 citations counted in INSPIRE as of 03 May 2018	

%\cite{Buttazzo:2013uya}
\bibitem{Buttazzo:2013uya} 
D.~Buttazzo, G.~Degrassi, P.~P.~Giardino, G.~F.~Giudice, F.~Sala, A.~Salvio and A.~Strumia,
``Investigating the near-criticality of the Higgs boson,''
JHEP {\bf 1312}, 089 (2013)
%doi:10.1007/JHEP12(2013)089
[arXiv:1307.3536 [hep-ph]].
%%CITATION = doi:10.1007/JHEP12(2013)089;%%
%772 citations counted in INSPIRE as of 03 May 2018

%\cite{Espinosa:2015qea}
%\bibitem{Espinosa:2015qea} 
%J.~R.~Espinosa, G.~F.~Giudice, E.~Morgante, A.~Riotto, L.~Senatore, A.~Strumia and N.~Tetradis,
%``The cosmological Higgstory of the vacuum instability,''
%JHEP {\bf 1509}, 174 (2015)
%doi:10.1007/JHEP09(2015)174
%[arXiv:1505.04825 [hep-ph]].
%%CITATION = doi:10.1007/JHEP09(2015)174;%%
%31 citations counted in INSPIRE as of 04 Jan 2016


	
\bibitem{Cheung:2007st} 
C.~Cheung, P.~Creminelli, A.~L.~Fitzpatrick, J.~Kaplan and L.~Senatore,
``The Effective Field Theory of Inflation,''
JHEP {\bf 0803}, 014 (2008)
% doi:10.1088/1126-6708/2008/03/014
[arXiv:0709.0293 [hep-th]].	

%\cite{Senatore:2010wk}
\bibitem{Senatore:2010wk} 
L.~Senatore and M.~Zaldarriaga,
``The Effective Field Theory of Multifield Inflation,''
JHEP {\bf 1204}, 024 (2012)
%doi:10.1007/JHEP04(2012)024
[arXiv:1009.2093 [hep-th]].
%%CITATION = doi:10.1007/JHEP04(2012)024;%%
%152 citations counted in INSPIRE as of 05 May 2018


%\cite{Kumar:2017ecc}
\bibitem{Kumar:2017ecc} 
S.~Kumar and R.~Sundrum,
``Heavy-Lifting of Gauge Theories By Cosmic Inflation,''
JHEP {\bf 1805}, 011 (2018)
%doi:10.1007/JHEP05(2018)011
[arXiv:1711.03988 [hep-ph]].
%%CITATION = doi:10.1007/JHEP05(2018)011;%%
%9 citations counted in INSPIRE as of 24 Sep 2018

%\cite{He:2018gyf}
\bibitem{He:2018gyf} 
  M.~He, A.~A.~Starobinsky and J.~Yokoyama,
  ``Inflation in the mixed Higgs-$R^2$ model,''
  JCAP {\bf 1805}, no. 05, 064 (2018)
  %doi:10.1088/1475-7516/2018/05/064
  [arXiv:1804.00409 [astro-ph.CO]].
  %%CITATION = doi:10.1088/1475-7516/2018/05/064;%%
  %12 citations counted in INSPIRE as of 17 Nov 2018

%\cite{Chen:2018xck}
\bibitem{Chen:2018xck} 
  X.~Chen, Y.~Wang and Z.~Z.~Xianyu,
  ``Neutrino Signatures in Primordial Non-Gaussianities,''
  JHEP {\bf 1809}, 022 (2018)
  %doi:10.1007/JHEP09(2018)022
  [arXiv:1805.02656 [hep-ph]].
  %%CITATION = doi:10.1007/JHEP09(2018)022;%%
  %2 citations counted in INSPIRE as of 16 Nov 2018
	
%\cite{Elliston:2012ab}
\bibitem{Elliston:2012ab} 
J.~Elliston, D.~Seery and R.~Tavakol,
``The inflationary bispectrum with curved field-space,''
JCAP {\bf 1211}, 060 (2012)
%doi:10.1088/1475-7516/2012/11/060
[arXiv:1208.6011 [astro-ph.CO]].
%%CITATION = doi:10.1088/1475-7516/2012/11/060;%%
%42 citations counted in INSPIRE as of 16 Apr 2018

\bibitem{Baumann:2011su} 
D.~Baumann and D.~Green,
``Equilateral Non-Gaussianity and New Physics on the Horizon,''
JCAP {\bf 1109}, 014 (2011)
%doi:10.1088/1475-7516/2011/09/014
[arXiv:1102.5343 [hep-th]].
	
\bibitem{Starobinsky:1986}
A.A. Starobinsky, Stochastic de sitter (inflationary) stage in the early universe, in Current topics in field theory, quantum gravity and strings, 
%in Current topics in field theory, quantum gravity and strings, 
H.J. de Vega and N. Sanchez eds., Springer (1986).

%\cite{Starobinsky:1994bd}
\bibitem{Starobinsky:1994bd} 
A.~A.~Starobinsky and J.~Yokoyama,
``Equilibrium state of a selfinteracting scalar field in the De Sitter background,''
Phys.\ Rev.\ D {\bf 50}, 6357 (1994)
%doi:10.1103/PhysRevD.50.6357
[astro-ph/9407016].
%%CITATION = doi:10.1103/PhysRevD.50.6357;%%
%313 citations counted in INSPIRE as of 17 Apr 2017


\bibitem{Achucarro:2010jv} 
A.~Achucarro, J.~O.~Gong, S.~Hardeman, G.~A.~Palma and S.~P.~Patil,
``Mass hierarchies and non-decoupling in multi-scalar field dynamics,''
Phys.\ Rev.\ D {\bf 84}, 043502 (2011)
%doi:10.1103/PhysRevD.84.043502
[arXiv:1005.3848 [hep-th]].

\bibitem{Achucarro:2012sm} 
A.~Achucarro, J.~O.~Gong, S.~Hardeman, G.~A.~Palma and S.~P.~Patil,
``Effective theories of single field inflation when heavy fields matter,''
JHEP {\bf 1205}, 066 (2012)
%doi:10.1007/JHEP05(2012)066
[arXiv:1201.6342 [hep-th]].

%\cite{Gong:2012ri}
\bibitem{Gong:2012ri} 
J.~O.~Gong, H.~M.~Lee and S.~K.~Kang,
%``Inflation and dark matter in two Higgs doublet models,''
JHEP {\bf 1204}, 128 (2012)
%doi:10.1007/JHEP04(2012)128
[arXiv:1202.0288 [hep-ph]].
%%CITATION = doi:10.1007/JHEP04(2012)128;%%
%32 citations counted in INSPIRE as of 12 Jan 2019

\bibitem{Weinberg:2005vy} 
S.~Weinberg,
``Quantum contributions to cosmological correlations,''
Phys.\ Rev.\ D {\bf 72}, 043514 (2005)
%doi:10.1103/PhysRevD.72.043514
[hep-th/0506236].

\bibitem{Chen:2010xka} 
X.~Chen,
``Primordial Non-Gaussianities from Inflation Models,''
Adv.\ Astron.\  {\bf 2010}, 638979 (2010)
%doi:10.1155/2010/638979
[arXiv:1002.1416 [astro-ph.CO]].

\bibitem{Wang:2013eqj} 
Y.~Wang,
``Inflation, Cosmic Perturbations and Non-Gaussianities,''
Commun.\ Theor.\ Phys.\  {\bf 62}, 109 (2014)
%doi:10.1088/0253-6102/62/1/19
[arXiv:1303.1523 [hep-th]].

\bibitem{Chen:2015dga} 
X.~Chen, M.~H.~Namjoo and Y.~Wang,
``On the equation-of-motion versus in-in approach in cosmological perturbation theory,''
JCAP {\bf 1601}, no. 01, 022 (2016)
%doi:10.1088/1475-7516/2016/01/022
[arXiv:1505.03955 [astro-ph.CO]].

%\cite{Chen:2012ge}
\bibitem{Chen:2012ge} 
X.~Chen and Y.~Wang,
``Quasi-Single Field Inflation with Large Mass,''
JCAP {\bf 1209}, 021 (2012)
%doi:10.1088/1475-7516/2012/09/021
[arXiv:1205.0160 [hep-th]].
%%CITATION = doi:10.1088/1475-7516/2012/09/021;%%
%71 citations counted in INSPIRE as of 09 Oct 2018

%\cite{Pi:2012gf}
\bibitem{Pi:2012gf} 
S.~Pi and M.~Sasaki,
%``Curvature Perturbation Spectrum in Two-field Inflation with a Turning Trajectory,''
JCAP {\bf 1210}, 051 (2012)
%doi:10.1088/1475-7516/2012/10/051
[arXiv:1205.0161 [hep-th]].
%%CITATION = doi:10.1088/1475-7516/2012/10/051;%%
%61 citations counted in INSPIRE as of 15 Mar 2019

%\cite{An:2017hlx}
\bibitem{An:2017hlx} 
H.~An, M.~McAneny, A.~K.~Ridgway and M.~B.~Wise,
``Quasi Single Field Inflation in the non-perturbative regime,''
JHEP {\bf 1806}, 105 (2018)
%doi:10.1007/JHEP06(2018)105
[arXiv:1706.09971 [hep-ph]].
%%CITATION = doi:10.1007/JHEP06(2018)105;%%
%12 citations counted in INSPIRE as of 26 Sep 2018

%\cite{Chen:2017ryl}
\bibitem{Chen:2017ryl} 
  X.~Chen, Y.~Wang and Z.~Z.~Xianyu,
  %``Schwinger-Keldysh Diagrammatics for Primordial Perturbations,''
  JCAP {\bf 1712}, no. 12, 006 (2017)
  doi:10.1088/1475-7516/2017/12/006
  [arXiv:1703.10166 [hep-th]].
  %%CITATION = doi:10.1088/1475-7516/2017/12/006;%%
  %15 citations counted in INSPIRE as of 26 Dec 2018

\bibitem{Gwyn:2012mw} 
R.~Gwyn, G.~A.~Palma, M.~Sakellariadou and S.~Sypsas,
``Effective field theory of weakly coupled inflationary models,''
JCAP {\bf 1304}, 004 (2013)
%doi:10.1088/1475-7516/2013/04/004
[arXiv:1210.3020 [hep-th]].

%\cite{Gwyn:2014doa}
\bibitem{Gwyn:2014doa} 
R.~Gwyn, G.~A.~Palma, M.~Sakellariadou and S.~Sypsas,
%``On degenerate models of cosmic inflation,''
JCAP {\bf 1410}, no. 10, 005 (2014)
%doi:10.1088/1475-7516/2014/10/005
[arXiv:1406.1947 [hep-th]].
%%CITATION = doi:10.1088/1475-7516/2014/10/005;%%
%13 citations counted in INSPIRE as of 15 Mar 2019

%\cite{Ashoorioon:2018uey}
\bibitem{Ashoorioon:2018uey} 
A.~Ashoorioon, R.~Casadio, M.~Cicoli, G.~Geshnizjani and H.~J.~Kim,
%``Extended Effective Field Theory of Inflation,''
JHEP {\bf 1802}, 172 (2018)
%doi:10.1007/JHEP02(2018)172
[arXiv:1802.03040 [hep-th]].
%%CITATION = doi:10.1007/JHEP02(2018)172;%%
%5 citations counted in INSPIRE as of 15 Mar 2019

%\cite{Tong:2017iat}
\bibitem{Tong:2017iat} 
X.~Tong, Y.~Wang and S.~Zhou,
``On the Effective Field Theory for Quasi-Single Field Inflation,''
JCAP {\bf 1711}, no. 11, 045 (2017)
%doi:10.1088/1475-7516/2017/11/045
[arXiv:1708.01709 [astro-ph.CO]].
%%CITATION = doi:10.1088/1475-7516/2017/11/045;%%
%4 citations counted in INSPIRE as of 14 Apr 2018

%\cite{Iyer:2017qzw}
\bibitem{Iyer:2017qzw} 
A.~V.~Iyer, S.~Pi, Y.~Wang, Z.~Wang and S.~Zhou,
``Strongly Coupled Quasi-Single Field Inflation,''
JCAP {\bf 1801}, no. 01, 041 (2018)
%doi:10.1088/1475-7516/2018/01/041
[arXiv:1710.03054 [hep-th]].
%%CITATION = doi:10.1088/1475-7516/2018/01/041;%%
%4 citations counted in INSPIRE as of 14 Apr 2018

%\cite{An:2017rwo}
\bibitem{An:2017rwo} 
  H.~An, M.~McAneny, A.~K.~Ridgway and M.~B.~Wise,
  %``Non-Gaussian Enhancements of Galactic Halo Correlations in Quasi-Single Field Inflation,''
  Phys.\ Rev.\ D {\bf 97}, no. 12, 123528 (2018)
  %doi:10.1103/PhysRevD.97.123528
  [arXiv:1711.02667 [hep-ph]].
  %%CITATION = doi:10.1103/PhysRevD.97.123528;%%
  %11 citations counted in INSPIRE as of 14 Dec 2018

%\cite{Ade:2015ava}
\bibitem{Ade:2015ava} 
  P.~A.~R.~Ade {\it et al.} [Planck Collaboration],
  %``Planck 2015 results. XVII. Constraints on primordial non-Gaussianity,''
  Astron.\ Astrophys.\  {\bf 594}, A17 (2016)
  %doi:10.1051/0004-6361/201525836
  [arXiv:1502.01592 [astro-ph.CO]].
  %%CITATION = doi:10.1051/0004-6361/201525836;%%
  %552 citations counted in INSPIRE as of 20 Dec 2018

\end{thebibliography}
\end{document}